\documentclass[twocolumn,preprintnumbers,amsmath,amssymb,aps]{revtex4-1} 
\usepackage{amsmath} 
\usepackage[utf8]{inputenc} 
\usepackage[english]{babel} 
\usepackage{graphicx} 
\usepackage{listings} 
\usepackage{soul}
\usepackage{color}

\usepackage{multirow}
\usepackage{lipsum}

\usepackage{pifont}
\usepackage[T1]{fontenc}
\usepackage{lmodern}

\begin{document}

\title{Different scenarios of dynamic coupling in glassy colloidal mixtures}

\author{Marco Laurati$^{a,b}$, Tatjana Sentjabrskaja$^{b}$, Jos\'e Ruiz-Franco$^{c}$, Stefan U. Egelhaaf$^{b}$, Emanuela Zaccarelli$^{c,d}$}

\address{$^{a}$Divisi\'on de Ciencias e Ingenier\'ias, Campus Le\'on, Universidad de Guanajuato, Loma del Bosque 103, Lomas del Campestre, 37150 Le\'{o}n, Guanajuato, Mexico}
\address{$^{b}$Condensed Matter Physics Laboratory, Heinrich Heine University, Universit{\"a}tsstr. 1, 40225 D{\"u}sseldorf, Germany}
\address{$^{c}$Dipartimento di Fisica, Universit\`a di Roma La Sapienza, Piazzale A. Moro 2, Roma 00185, Italy}
\address{$^{d}$CNR-ISC, Universit\`a di Roma La Sapienza, Piazzale A. Moro 2, Roma 00185, Italy}
\date{\today}

\begin{abstract}

Colloidal mixtures represent a versatile model system to study transport in complex environments. They allow for a systematic variation of the control parameters, namely size ratio, total volume fraction and composition. We study the effects of these parameters on the dynamics of dense suspensions using molecular dynamics simulations and differential dynamic microscopy experiments. We investigate the motion of the small particles through the matrix of large particles as well as the motion of the large particles. A particular focus is on the coupling of the collective dynamics of the small and large particles and on the different mechanisms leading to this coupling. For large size ratios, about 1:5, and an increasing fraction of small particles, the dynamics of the two species become increasingly coupled and reflect the structure of the large particles. This is attributed to the dominant effect of the large particles on the motion of the small particles which is mediated by the increasing crowding of the small particles. Furthermore, for moderate size ratios, about 1:3, and sufficiently high fractions of small particles, mixed cages are formed and hence the dynamics are also strongly coupled. Again, the coupling becomes weaker as the fraction of small particles is decreased. In this case, however, the collective intermediate scattering function of the small particles shows a logarithmic decay corresponding to a broad range of relaxation times.

\end{abstract}

\maketitle

\section{Introduction}

Several materials of everyday use are found in an amorphous solid state, among them window glass, plastics, ceramics, foodstuff \cite{larson1999}.  Glassy materials of commercial use are often formed by either particles with a certain size distribution or even more frequently by several components presenting largely different sizes and mobilities \cite{shelby}. Often, studies on the dynamical properties of these materials concentrate on properties that are averaged over the different sizes and species\cite{brambilla,commentbrambillabill}. However, studies on colloidal model systems of spherical particles have shown that dynamics in the presence of polydispersity or of multiple components can be highly heterogeneous, with the dynamic heterogeneity strongly linked to the presence of different particle sizes \cite{imhof,emanuela_poly,heckendorf,tanja_sm2014,tanja_review,Moreno_jcp,Moreno_jcp2,mayer2008asymmetric,Mayer_zaccarelli,narumi2011,hamanaka2007}. 

Binary colloidal mixtures of hard spheres have been used as the simplest model for multi-component systems. Studies on binary glasses of hard spheres revealed that several dynamical behaviors are observed especially when the size difference between the two species is large \cite{Thomas_EPL,maldonado,imhof,williams2001,tanja_review,germain2009,yunker2010,lynch2008,han2018}. For instance for size ratios $\delta = R_\mathrm{s}/R_\mathrm{L} = 0.1$ and $0.2$, where $R_\mathrm{s}$ and $R_\mathrm{L}$ are the radii of the small and large particles respectively, several glass and even gel states are observed, in which either both species are dynamically arrested, or the small component remains mobile within the glass of the large one \cite{Thomas_EPL,maldonado,imhof,hendricks}. In a recent work\cite{nat_comm} we have investigated the dependence of the dynamics of the small particles on the size ratio in the limit of a very small fraction of small particles. This limit is particularly interesting because the small particles, due to their low concentration, can be considered to be non-interacting among themselves. This resembles the idealized Lorentz gas model\cite{Lorentz} in a realistic situation.  It was shown that at small $\delta$ a diffusive behavior is observed, whereas at a critical $\delta \approx 0.3$ the dynamics become anomalous. In simulations and experiments a logarithmic decay was observed in the intermediate scattering functions of the small particles. This anomalous dynamics was found to be strongly related to the slow dynamics of the large component, in contrast to the Lorentz model and models which consider the motion of particles through an immobile porous matrix\cite{Hofling_PhysRep,krackoviak}. 

In the present study we investigate a broad range of binary mixtures with different size ratios, total packing fractions and mixture compositions, thus extending a previous study that explored the limit of dilute small particles \cite{nat_comm}. In particular, we investigate the coupling of the long-time dynamics of the large and small spheres using simulations and experiments. The mixture composition is quantified by the mixing ratio $x_\mathrm{s} = \phi_\mathrm{s}/\phi$, with $\phi_\mathrm{s}$ and $\phi$ the volume fractions of the small and all particles, respectively. We show that the anomalous behavior observed for small $x_\mathrm{s}$ and a critical $\delta$ disappears upon increasing the fraction of small particles, while a coupling between the long-time dynamics of large and small particles becomes more and more evident. In addition, we find different scenarios for the glassy dynamics of the mixture depending on whether the small particles are trapped in the voids of the large particles, or whether the two species form joint cages. Since a binary mixture is the simplest case of a multi-component system, these results might have implications for the understanding of glasses formed by several components and, more general, the motion of small particles in slowly-rearranging crowded environments.
We perform experiments using Differential Dynamic Microscopy (DDM)\cite{cerbino_prl,cerbino_dfm,cerbino2017perspective}. Although this technique is based on microscopy, it provides information analogous to that typically obtained from dynamic light scattering (DLS) experiments\cite{berne-pecora}, namely the intermediate scattering function $f(q,\Delta t)$ as a function of delay time $\Delta t$, with $q$ the scattering vector which determines the length scale on which the particle dynamics are probed. We exploit the advantage of DDM that fluctuations of the incoherent fluorescence signal can be analyzed, a possibility which is excluded in DLS, that requires coherent light. Furthermore, the use of a confocal microscope reduces the effect of background fluorescence, significantly improving the determination of $f(q,\Delta t)$ compared to a conventional fluorescence microscope. Thus DDM allows us to obtain information on a single species, here the small particles, in a multi-component sample.

\section{Materials and Methods}

\subsection{Experiments}

{\bf Materials.} We studied dispersions of sterically
stabilized polymethylmethacrylate (PMMA) particles of diameters $\sigma_{\mathrm{l}^{(1)}}= 3.10$~$\mu$m (polydispersity 0.07, not fluorescently labeled) or $\sigma_{\mathrm{l}^{(2)}} = 1.98$~$\mu$m (polydispersity 0.07, not fluorescently labeled) mixed with particles of diameter $\sigma_{\mathrm{s}} = 0.56$~$\mu$m (polydispersity 0.13, fluorescently labeled with nitrobenzoxadiazole (NBD)),
in a cis-decalin/cycloheptyl-bromide  mixture 
that closely matches the density and refractive index of the particles. Therefore the size ratio of the mixtures was $\delta = 0.18$ and  $\delta = 0.28$, respectively. 
With added salt (tetrabutylammoniumchloride), this system presents hard-sphere like
interactions \cite{yethiraj03,Poon/Weeks/Royall}. 
The volume fraction of a sedimented sample of the large particles was estimated to be $\phi = 0.65$ by comparing with numerical simulations and experiments \cite{silescu, weeks}, where the uncertainty $\Delta\phi$ is typically 3$~$\%, possibly larger\cite{poon2012}.
We used this volume fraction $\phi$ of the large particles as a reference value, while the volume fraction of the stock suspension containing the small particles was adjusted in order to obtain comparable linear viscoelastic moduli. 
Due to the different sizes of the particles, the trivial dependence of the moduli on the particle size was accounted for by a normalisation with the energy density $3k_{\mathrm{B}}T/4\pi R^3$, where $k_{\mathrm{B}}$ is the Boltzmann constant and $T$ the temperature. Similarly, the frequency was normalized by the free-diffusion Brownian time $t_0 = 6 \pi \eta R^3/k_{\mathrm{B}}T$, where $\eta=$ 2.2~mPa$\,$s is the solvent viscosity. Following this procedure we obtained stock suspensions of large and small particles with comparable rheological properties and, following the generalised Stokes-Einstein relation \cite{mason2000}, also comparable dynamics. Given this assumption of dynamical equivalence,  the stock suspensions of large and small particles are expected to have a similar location with respect to the glass transition. Thus variations in the dynamical arrest of the mixtures can be exclusively attributed to the composition which enables us to study the effects of mixing on the glass behaviour. It is important to note that this would not be possible by mixing stock suspensions with identical volume fraction $\phi$ since, due to the different polydispersities, the volume fractions at the glass transition are different. Accordingly, the comparable dynamics but the different polydispersities imply slightly different volume fractions of the two stock suspensions.
Samples with different total volume fractions and different composition, namely a fraction of small particles $x_{\mathrm{s}} = 0.01$ and  0.05,  were prepared by mixing the one-component stock suspensions and subsequent dilution.\\ 
 
{\bf DDM measurements.} We acquired confocal microscopy images in a
plane at a depth of approximately 30~$\mu$m from the coverslip.  Images with 512$\times$512 pixels,
corresponding to 107~$\mu$m $\times$ 107~$\mu$m, were acquired at two different rates: a fast rate of 30 frames per second to follow the short-time dynamics and a slow rate,
 between 0.07 and 0.33  frames per second, depending on sample, to follow the long-time dynamics. A Nikon A1R-MP confocal scanning unit mounted on a Nikon Ti-U inverted microscope, equipped with a 60x Nikon Plan Apo oil immersion objective (NA = 1.40), was used to capture image series. The pixel size corresponding to this magnification is 0.21~$\mu$m $\times$ 0.21~$\mu$m. The confocal images were acquired using the maximum pinhole aperture, corresponding to a pinhole diameter of 255~$\mu$m.   
Time series of 10$^4$ images were acquired for 2 to 5 distinct volumes, depending on sample.\\

We call $i(x,y,t)$ the intensity measured at time $t$ in a pixel with coordinates $x$ and $y$. 
The difference between two intensity patterns separated by a delay time $\Delta t$ is calculated, $\Delta i(x,y,t+\Delta t) = i(x,y,t+\Delta t)- i(x,y,t)$, and subsequently Fourier transformed to yield $\Delta \hat{i}({\bf q},t,\Delta t)$, where ${\bf q} =(q_x,q_y)$\cite{lu2012}. For a stationary signal and isotropic scattering this is related to the structure function $D(q,\Delta t)$:\cite{cerbino_prl,cerbino2017perspective}
\begin{equation}
D(q,\Delta t) = \langle |\Delta \hat{i}(q,t,\Delta t)|^2\rangle
\label{eq_diqt_dfm}
\end{equation}
where the notation $\langle \rangle$ represents a time and ensemble average. This is related to the intermediate scattering function $f(q,\Delta t)$  by:
\begin{equation}
D(q,\Delta t) = A(q)[1-f(q,\Delta t)]+B(q)
\label{eq_diqt_fqt}
\end{equation}
where $A(q)= N|\hat{K}(q)|^2S(q)$, $N$ is the particle number in the observed volume, $\hat{K}(q)$ is the Fourier transform of the point-spread function of the microscope, $S(q)$ is the static structure factor of the system, and $B(q)$ contains the information related to the background noise, in particular the camera noise.

\subsection{Simulations}


Event-driven Molecular Dynamics simulations were performed in the $NVT$ ensemble. We simulate binary mixtures of hard particles of mass $m$ in a cubic box of size $L$. We consider monodisperse small particles of diameter $\sigma_\mathrm{s}$ and polydisperse large particles with average diameter $\sigma_\mathrm{l}$ and polydispersity $0.07$ in order to avoid crystallization\cite{Zacca09}. We vary the size ratio in the range $0.2\leq\delta\leq0.5$, the total volume fraction $0.55 \leq \phi \leq 0.64$,  and investigate three values of the mixing ratio, $x_\mathrm{s}=0.01, 0.05$ and $0.10$. We ensure that the number of small particles $N_\mathrm{S}$ is always larger than $250$ and thus we have total number of particles  $2000 \lesssim N \lesssim 5000$. Mass, length and energy are measured in units of $m$, $\sigma_\mathrm{l}$, and $k_BT$.  Simulation time is in units of $t_0=\sqrt{m \sigma_\mathrm{l}^2/k_B T}$ and we fixed $k_{\mathrm{B}}T=1$. After equilibration of the system, simulations were performed for a maximum total time of $\sim 10^5 t_0$ for a maximum of three realizations for each state point, particularly those at high $\phi$. Thermodynamic and dynamic observables have been calculated performing time and ensemble averages.

We calculate the partial static structure factors for large and small particles as:
\begin{equation}
	\label{eq:Sq}
	S_{jj}(q)=\frac{1}{N_j}\left\langle \sum_{\alpha,\beta}^{N_j} e^{-i\mathbf{q}\left(\mathbf{r}^j_{\alpha}-\mathbf{r}^j_{\beta}\right)}\right\rangle
\end{equation}
where $N_j$ is the number of particles of the species $j=L,S$ and $\mathbf{r}^{j}_{\alpha}$ is the corresponding position vector of particle $\alpha$. 
Similarly, we also calculate the partial intermediate scattering functions, defined as
\begin{equation}
	\label{eq:collective}
	f^j\left(q,\Delta t\right)=\left\langle \frac{1}{N_j}\sum_{\alpha,\beta}e^{-i\mathbf{q}\left(\mathbf{r}^j_{\alpha}\left(t\right)-\mathbf{r}^j_{\beta}\left(0\right)\right)} \right\rangle
\end{equation}
for large and small particles, respectively.
To monitor the self-dynamics of the particles, we have also calculated the 
probability $P^j(r, \Delta t)$ that a particle of species $j$ has moved a distance $r$ in an interval of time $\Delta t$. This is defined as,
\begin{equation}
	\label{eq:self_vanHove}
	P^j(r,\Delta t) = 4\pi r^2 G_{s}^{j}\left(r,\Delta t\right)
\end{equation}
where $G_s^j(r,\Delta t)$ is the self part of the van Hove correlation function\cite{hansenbook}.


\begin{figure} [ !tb]
      \includegraphics[angle=0, width=0.47\textwidth] {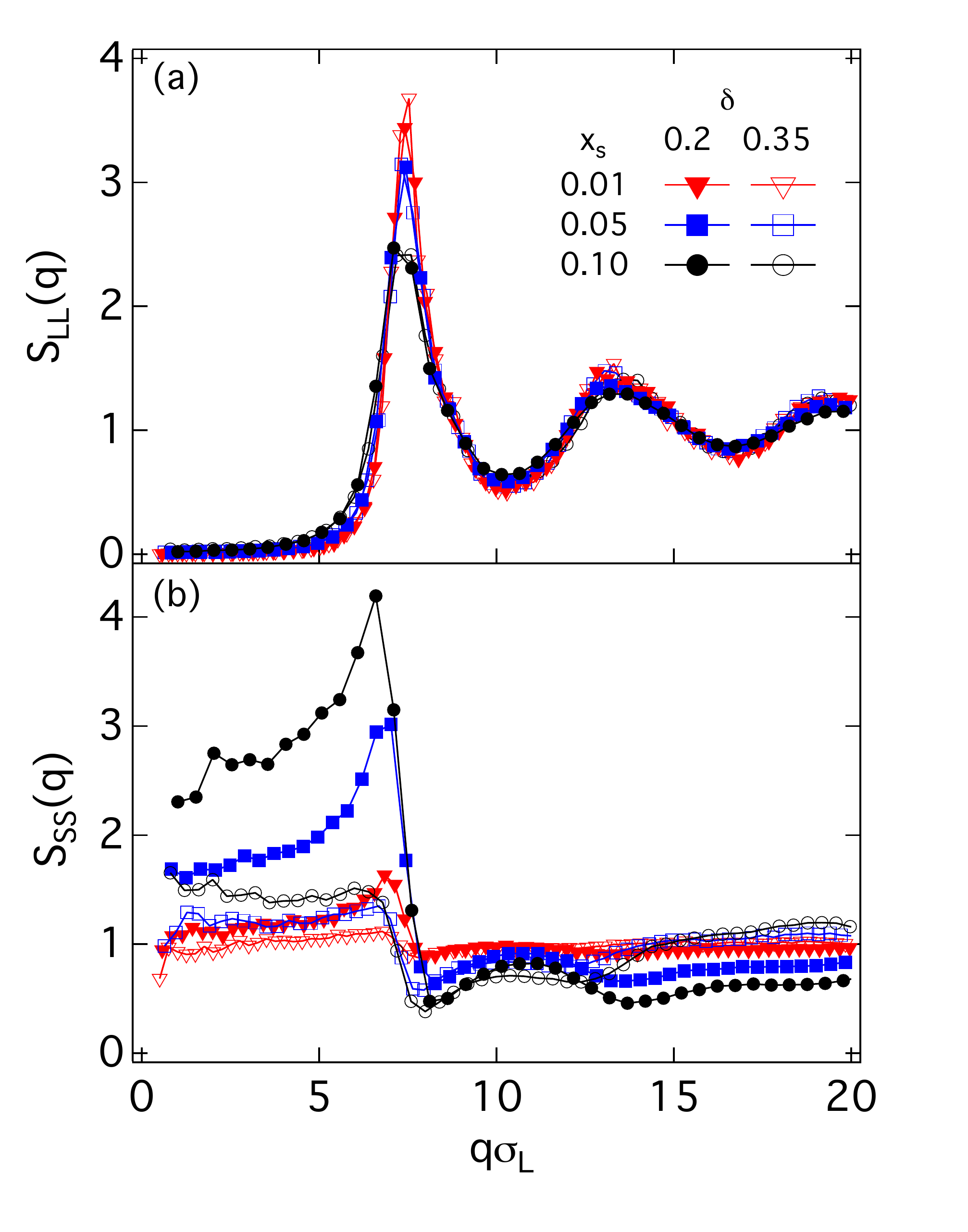}
             \caption{Static structure factors of (a) the large  particles, $S_\mathrm{LL}(q)$, and (b) the small particles, $S_\mathrm{SS}(q)$, from simulations for size ratios $\delta = 0.2$ (full symbols) and 0.35 (open symbols), total volume fraction $\phi = 0.62$ and different values of the fraction of small particles, $x_\mathrm{s}$ (as indicated).}  
\label{sq_comp}         
\end{figure}

\section{Results and Discussion}

\subsection{Structure}

In this section we report results on the structural organization of small and large particles obtained from simulations. These results will serve as a guide for the interpretation of the dynamics at different length scales, which is presented in the following sections. Experimental structural data are not available. The large particles are not fluorescently labeled and hence not visible in the confocal microscope. For the small particles, particle tracking was not possible because they are too small to be accurately located. Also DDM could not be used to obtain the partial structure factors of the small spheres through the parameter $A(q)$ in Eq. 2 \cite{lu2012,nat_comm}. This approach is based on the measurement of a dilute sample for which the partial structure factor is known to be unity and from which hence the contribution of the point spread function can  be determined. However, the measurements of the dilute small particles and the concentrated mixtures, respectively, require different pinhole sizes resulting in different and unknown point spread functions. Hence this approach is not possible in the present situation.\\

We focus on glassy samples with $\phi = 0.62$ and different compositions ($x_\mathrm{s} = 0.01$, 0.05 and 0.10) and size ratios ($\delta = 0.2$ and 0.35). Data for different volume fractions ($\phi = 0.56$, 0.58 and 0.60), showing qualitatively similar trends, can be found in the supplemental material. The partial static structure factors of the large particles, $S_\mathrm{LL}(q)$, for all compositions and size ratios show  a disordered organization with pronounced peaks, which indicate strong correlations particularly for the smallest values of $x_\mathrm{s}$ (Fig.\ref{sq_comp}a). The main peak is located at $q\sigma_\mathrm{L} \approx 7.4$. Upon increasing the relative amount of small particles,  the peaks of $S_\mathrm{LL}(q)$ decrease in height and move to slightly smaller $q\sigma_\mathrm{L}$ values, as a result of the progressive dilution of the large particles. The differences between the $S_\mathrm{LL}(q)$ obtained for the two size ratios are minor. As suggested in a previous study, this could be related to the fact that for both size ratios the small particles can occupy the voids between the  large particles\cite{tanja_sm2013}.\\
The size ratio, however, affects the local structure of the small particles (Fig.\ref{sq_comp}b). For $\delta = 0.35$ the structure factor $S_\mathrm{SS}(q)$ indicate weak correlations, which grow only moderately with increasing $x_\mathrm{s}$. On the other hand, for $\delta = 0.2$ the correlations are considerably more pronounced and grow rapidly with increasing $x_\mathrm{s}$. In particular the low-$q$ value significantly increases with $x_\mathrm{s}$, indicating a much larger compressibility and deviations from an ideal behavior at large distances.  For both size ratios the main correlation peak, which is encountered at $q\sigma_\mathrm{L} \approx 6.7$ for  $x_\mathrm{s} = 0.01$, moves to slightly lower $q\sigma_\mathrm{L}$ values with increasing $x_\mathrm{s}$.  Thus the most-likely distance between the small particles becomes better defined and increases with $x_\mathrm{s}$, although the increased number of small particles would naively suggest its decrease. This may be due to the simultaneous dilution of the matrix of large particles, occurring by increasing $x_\mathrm{s}$ at constant $\phi$, which then allows the small particles to distribute more homogeneously within the available free volume, until their distribution is affected by other small particles at higher $x_\mathrm{s}$. 
We additionally note that the second peaks become more pronounced with $x_\mathrm{s}$ but barely depend on $\delta$. These correlations are found to slightly grow with increasing $x_\mathrm{s}$, suggesting  that the small particles pack more densely at the local scale, thus concentrating in the largest voids in between the large particles. For $\delta=0.5$, mixing effects are even more evident as reported in the supplemental material.

\subsection{Dynamics}

In this section we present results for the dynamics of the large and small particles, particularly examining the degree of coupling between the two species in glassy states.  Therefore, again we focus on a single volume fraction ($\phi = 0.62$) at different compositions ($x_\mathrm{s} = 0.01, 0.05, 0.10$) and different size ratios, $\delta = 0.2$ to $\delta = 0.5$. Data  for additional volume fractions are reported in the supplemental material. We characterize the collective dynamics of large and small particles using the intermediate scattering function.
First, we choose a fixed value of the scattering vector, $q\sigma_\mathrm{L} \approx 3.5$, where the most interesting effects, in particular anomalous logarithmic relaxations\cite{nat_comm}, are observed. After a qualitative discussion of the relation between the dynamics of the large and small particles, based on a direct comparison of the intermediate scattering functions, we discuss the relaxation times and plateau heights obtained by modelling the simulated and experimental intermediate scattering functions, which provide more quantitative evidence of coupling effects.

\subsubsection{Collective Intermediate Scattering Functions: Qualitative Observations\\ }
\begin{figure} [!tb]
\centering
     \includegraphics[angle=0, width=0.47\textwidth] {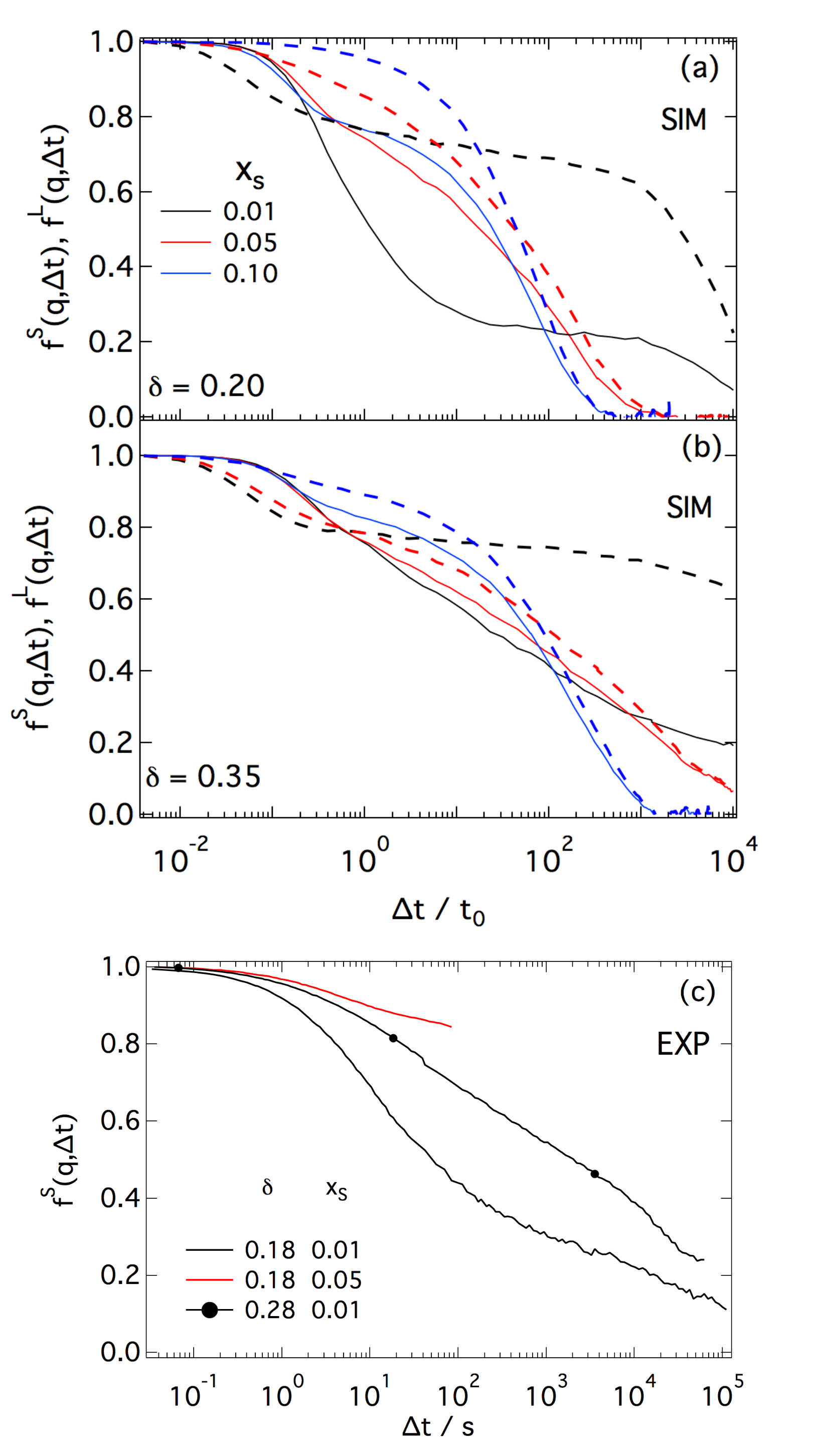}
\caption{(a,b) Intermediate scattering function $f^i(q,\Delta t)$ from simulations for $q\sigma_l = 3.5$, $\phi = 0.62$, $x_\mathrm{s} = 0.01$ (black), 0.05 (red), 0.1 (blue) and (a)  $\delta = 0.20$ (b) $\delta = 0.35$. Solid and dashed lines represent data of small and large particles, respectively. (c) Intermediate scattering function $f^{\mathrm{S}}(q,\Delta t)$ of the small particles from experiments for $q\sigma_l = 3.5$, $\phi = 0.61$ and indicated values of $x_\mathrm{s}$ and $\delta$.} 
\label{xs_dep}         
\end{figure}

We start by considering the size ratio $\delta = 0.2$ (Fig.\ref{xs_dep}a). For $x_\mathrm{s} = 0.01$ the intermediate scattering function of the large particles $f^L(q,\Delta t)$, obtained from simulations, shows a two-step decay, which is characteristic of glassy dynamics, with a long plateau at intermediate times that indicates a transient trapping of the particles into cages (Fig. \ref{xs_dep}a). The corresponding intermediate scattering functions of small particles $f^S(q,\Delta t)$ show a different behaviour, which is characterised by a pronounced initial decay followed by a plateau whose height is approximately a quarter of the one of the large particles. This indicates that  only a small fraction of the small particles is trapped in the matrix of large particles, while the majority of them is diffusing. Despite the difference in the height of the plateaus, both $f^L(q,\Delta t)$ and $f^S(q,\Delta t)$ seem to finally relax at a comparable time $\Delta t\approx 10^3 t_0$. This suggests that some small particles cannot escape the local confinement without the large particles moving and are thus slaved to the dynamics of the slower large particles. Experiments, in which only the dynamics of the small particles is available, show a comparable behavior of $f^S(q,\Delta t)$ (Fig.\ref{xs_dep}c).
At $x_\mathrm{s}=0.05$ (Fig. \ref{xs_dep}a), the decay of the $f^L(q,\Delta t)$ barely shows a two-step decay and the final relaxation occurs at shorter times. This can be attributed to the small concentration of large particles and the intercalation of small particles between large particles, which progressively disrupts the cage of large particles \cite{tanja_sm2013,tanja_sm2014,tanja_review}. The $f^S(q,\Delta t)$ correspondingly shows a faster final decay, while the intermediate plateau is much shorter and at a higher level. Once more the final decay of the two species is coupled. In contrast, the initial decay of $f^S(q,\Delta t)$ is independent of $x_\mathrm{s}$ indicating that the short-time motion of the small particles is independent of the large particles. The same qualitative behavior is observed in the experimental data, even though the restricted time window does not allow to observe the final decay of $f^S(q,\Delta t)$ (Fig.\ref{xs_dep}c). 
For an even larger fraction of small particles, $x_\mathrm{s} = 0.10$, the glass melting of the large particles becomes even more evident, with the $f^L(q,\Delta t)$ showing a decay which is almost exponential. At the same time the intermediate scattering function of the small particles shows a clear two-step decay, which can be associated to the formation of cages of small particles, as expected for the larger concentration of small particles. However also in this case the final relaxation appears to be coupled.

For the more moderate size disparity $\delta = 0.35$  and $x_\mathrm{s} = 0.01$ the $f^L(q,\Delta t)$ at short and intermediate times is similar to that of $\delta = 0.2$. The final decay though is apparently slower, indicating a larger degree of dynamical arrest (Fig.\ref{xs_dep}b). The $f^S(q,\Delta t)$ is completely different and shows a logarithmic decay. This behavior has been linked to an extremely broad distribution of relaxation times and the slow dynamics of the large particles \cite{nat_comm}. The same qualitative behaviour is observed in the experimental $f^S(q,\Delta t)$ (Fig.\ref{xs_dep}c). For this particular case thus, the dynamic coupling between the two species is much weaker. When increasing $x_\mathrm{s}$ to 0.05 the final decay of the large particles occurs at shorter times, similar to $\delta = 0.2$. However the acceleration of the dynamics is less pronounced than for $\delta = 0.20$ and a two-step decay is visible. The logarithmic decay of $f^S(q,\Delta t)$ is less evident and the final relaxation now follows again that of the large particles. For $x_\mathrm{s} = 0.10$ both species show a two-step decay and not only the final relaxation are similar, but also the intermediate plateaus. This suggests that if the particle sizes and volume fractions of the two species are more similar, mixed caging is favoured.

\begin{figure} [!tb]
\centering
     \includegraphics[angle=0, width=0.47\textwidth] {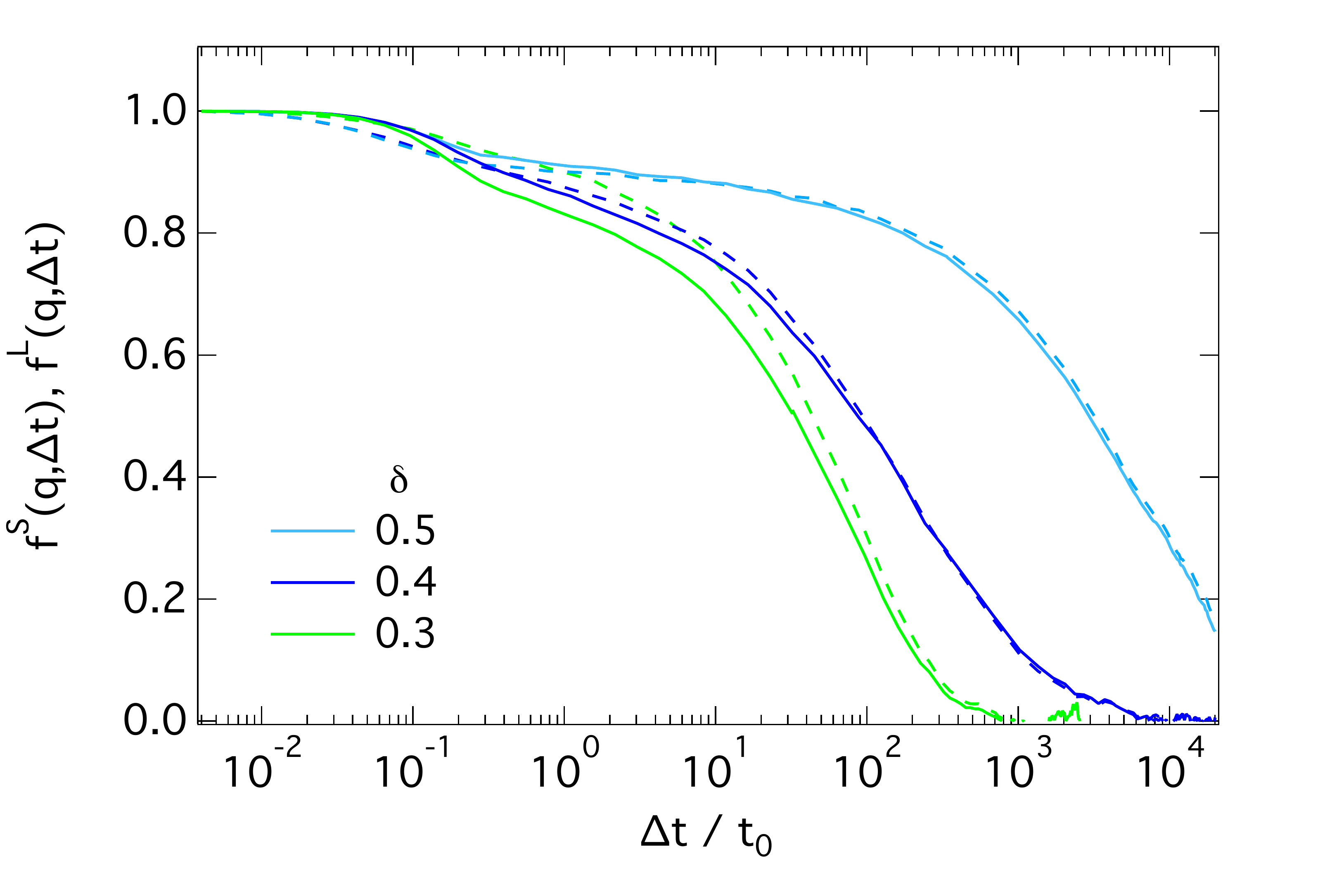}
\caption{Intermediate scattering functions $f^i(q,\Delta t)$ from simulations for $q\sigma_l = 3.5$, $\phi = 0.62$, $x_\mathrm{s} = 0.10$ and different values of $\delta$ (as indicated). Solid and dashed lines represent data of the small and large particles, respectively.}         
\label{fig:sizeratio}
\end{figure}
The behavior of the intermediate scattering functions (Fig. \ref{xs_dep}) suggests that the degree of dynamic coupling increases with $\delta$. To better quantify the dependence on $\delta$, simulations for $\phi = 0.62$, $x_\mathrm{s} = 0.10$  and size ratios $\delta = 0.3$, 0.4 and 0.5 were also performed and the corresponding intermediate scattering functions are reported in Fig.\ref{fig:sizeratio}. Data for other $x_\mathrm{s}$ can be found in the supplemental material. For $\delta = 0.4$ and even more pronounced for 0.5, the intermediate scattering functions of small and large particles almost overlap. This supports the conclusion that moderate size differences favour mixed caging and hence a strong coupling of the dynamics.

\begin{figure} [!htb]
   \centering
      \includegraphics[angle=0, width=0.45\textwidth] {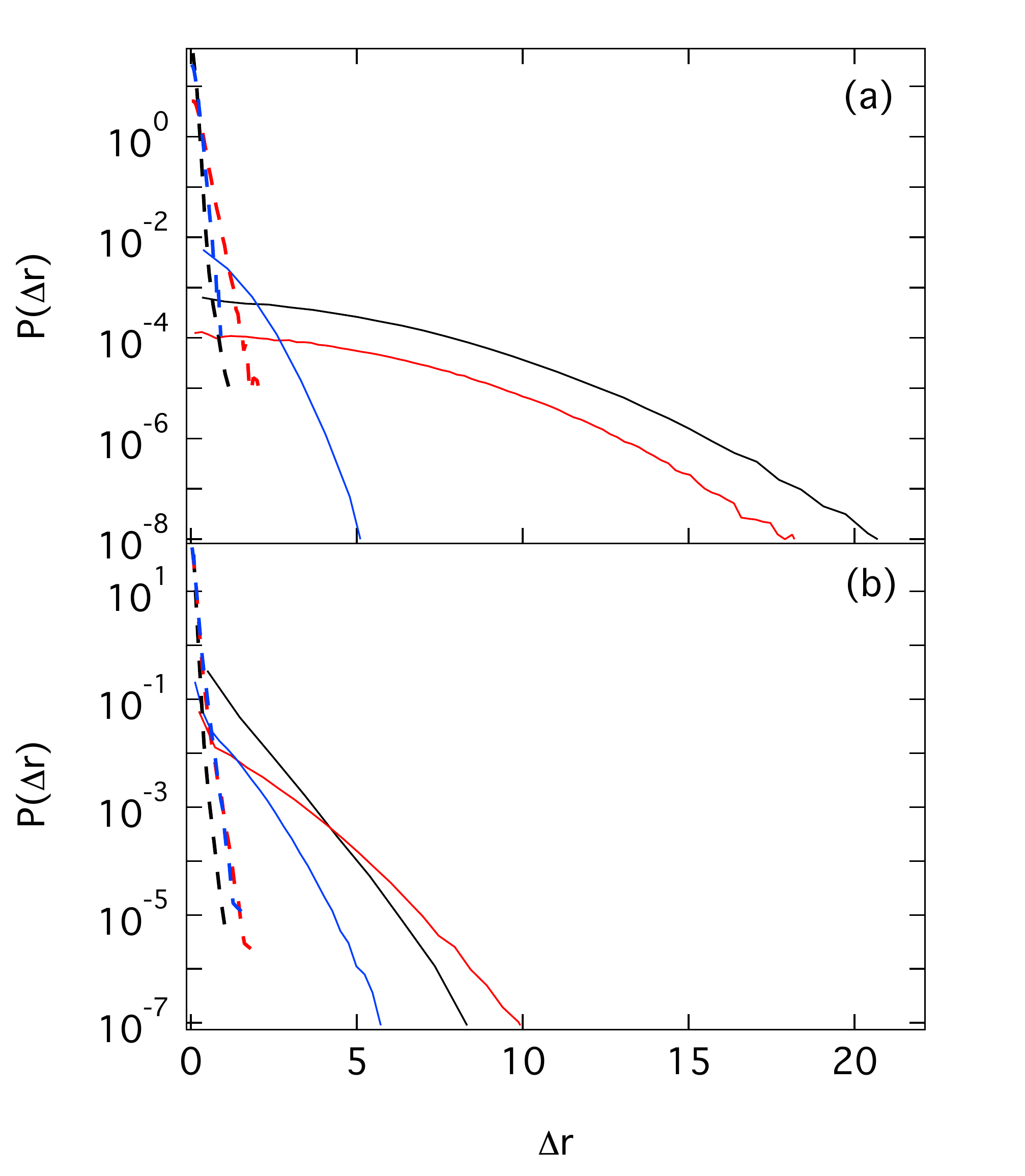}
         \caption{Displacement probability distribution $P(\Delta r)$ from simulations for $\Delta t = 338 t_0$, $\phi = 0.62$ and $x_\mathrm{s} = 0.01$ (black), 0.05 (red), 0.1 (blue). (a) $\delta = 0.20$, (b) $\delta = 0.35$. Solid and dashed lines represent data of the small and large particles, respectively.}   
\label{vh_xsdep}         
\end{figure}

Mode coupling theory predicts that the self-dynamics of small particles, for large size ratios, is completely decoupled from the behavior of the large particles\cite{bosse1995self}. This is found also in our simulations, as shown in Fig.\ref{vh_xsdep} where the displacement probability distributions $P(\Delta r)$  of both species are reported for a relatively long time ($\Delta t = 338 t_0$). The displacements of the small particles are considerably larger than those of the large particles for all $x_\mathrm{s}$ and $\delta$ considered. Interestingly, for $\delta = 0.35$ the width of the distribution displays a reentrance with mixture composition: it is broader for $x_\mathrm{s} = 0.05$ than for 0.01 and 0.10. This effect is particularly pronounced for the small particles but is also observed for the large particles. These results confirm that the dynamic coupling of the two species at long times is a feature encountered only in the collective dynamics.  


%

%

\subsection{Collective Intermediate Scattering Functions: Quantitative Analysis} 

\begin{figure} [!tb]
\centering
     \includegraphics[angle=0, width=0.45\textwidth] {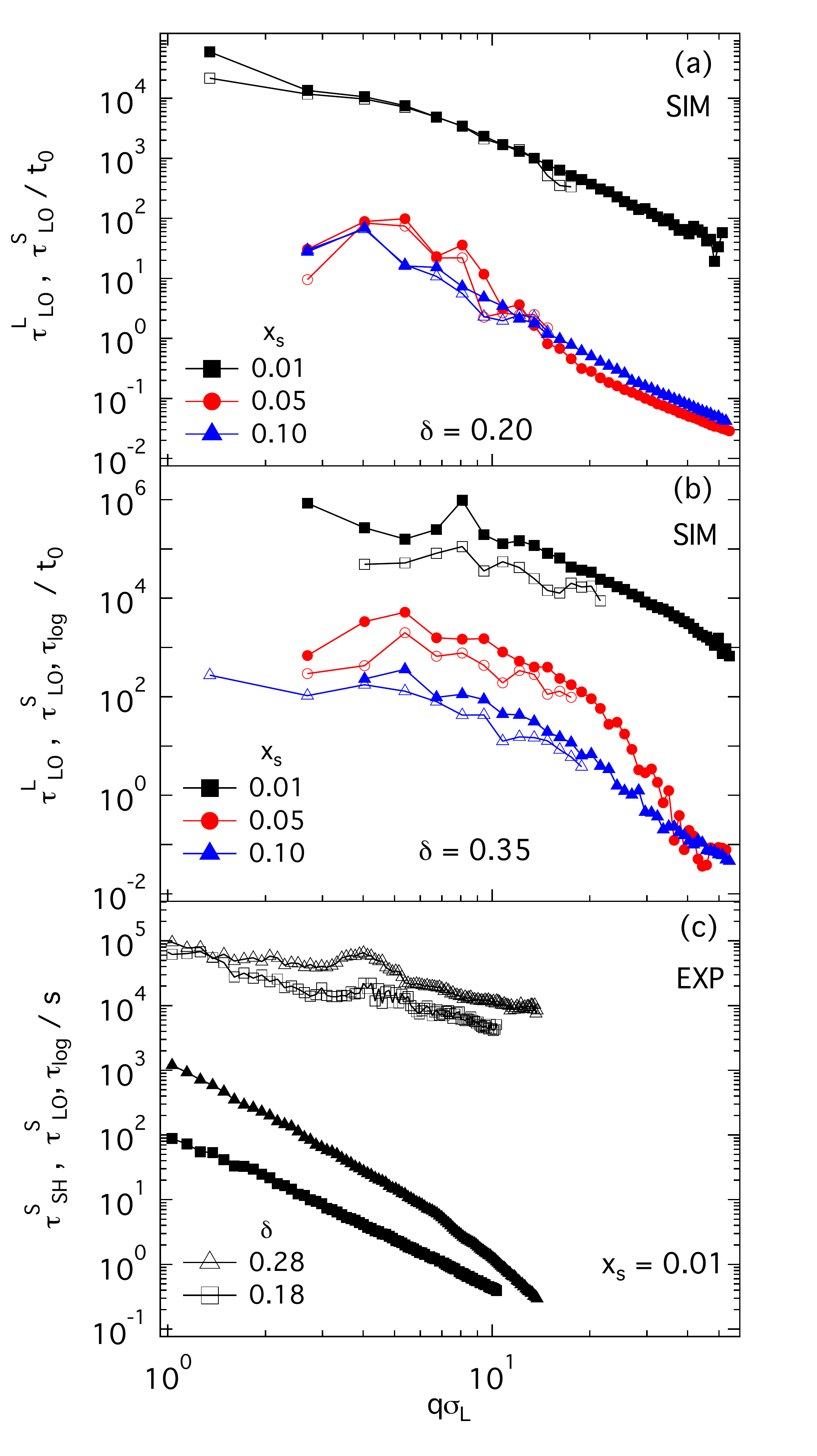}
\caption{(a,b) Relaxation times $\tau^j_\mathrm{LO}$ and $\tau_\mathrm{log}$ obtained by fitting Eq. \ref{stretch_exp_fit} or \ref{log_fit} to $f^{j}(q,\Delta t)$, obtained from simulations for $\phi = 0.62$, $x_\mathrm{s} = 0.01$ ($\Box$), 0.05 (\textcolor{red}{\Large $\circ$}), 0.10 (\textcolor{blue}{$\triangle$}) and (a) $\delta = 0.20$, (b) $\delta = 0.35$. Full symbols: Large particles;  Open Symbols: Small particles. (c) Relaxation times $\tau^S$ and $\tau_\mathrm{log}$  for the small particles from experiments, for $\phi =0.61$, $x_\mathrm{s} = 0.01$ and $\delta = 0.18$ ($\Box$) and $\delta = 0.28$ ($\triangle$). Open symbols: Long-time relaxation times $\tau^S_\mathrm{LO}$. Full symbols: Short time relaxation times $\tau^S_\mathrm{SH}$.}
\label{fit_params_1}         
\end{figure}

\begin{figure} [!tb]
\centering
     \includegraphics[angle=0, width=0.45\textwidth] {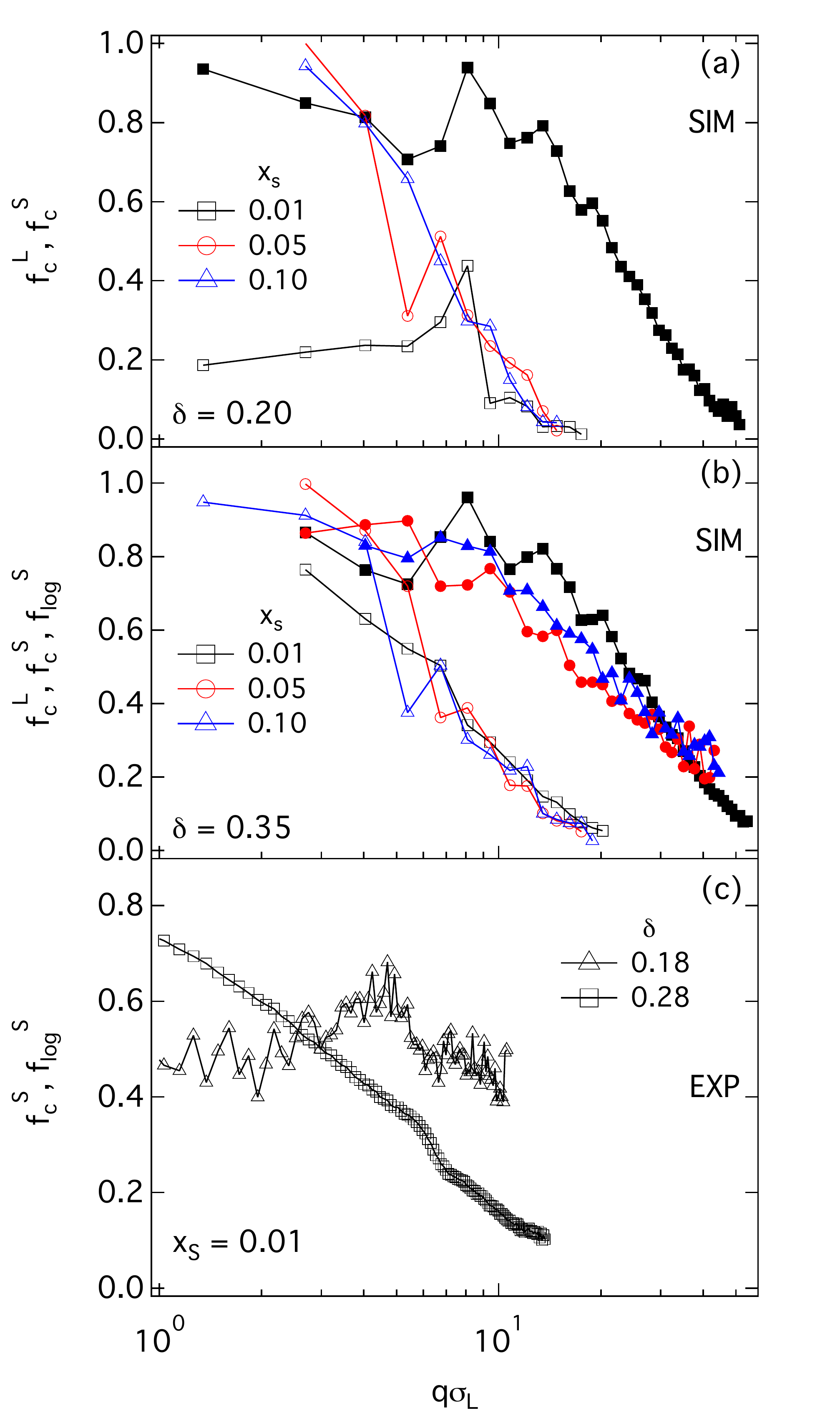}
\caption{(a,b) Plateau heights $f_c$ obtained by fitting Eq. \ref{stretch_exp_fit} or \ref{log_fit} to $f(q,\Delta t)$ from simulations for $\phi = 0.62$, $x_\mathrm{s} = 0.01$ ($\Box$), 0.05 (\textcolor{red}{\Large $\circ$}), 0.10 (\textcolor{blue}{$\triangle$}) and (a) $\delta = 0.20$, (b) $\delta = 0.35$. Full symbols: Large particles;  Open symbols: Small particles. (c) Plateau heights $f_c$ for the small particles from experiments for $\phi =0.61$, $x_\mathrm{s} = 0.01$ and $\delta = 0.18$ ($\Box$) and $\delta = 0.28$ ($\triangle$).}         
\label{fit_params_2} 
\end{figure}

To quantify the degree of coupling of the long-time collective dynamics we have modeled the long-time decay of $f(q,\Delta t)$ with a stretched exponential function:
\begin{equation}
f^{j}(q,\Delta t)=f_{c}^{j}\exp[-(\Delta t/\tau^j_\mathrm{LO})^{\beta^{j}}]
\label{stretch_exp_fit}
\end{equation}
where $f_{c}^{j}$ is the plateau height, $\tau^j_\mathrm{LO}$ the long-time relaxation time and $\beta^j$ the stretching exponent for $j = L, S$ indicating large and small particles, respectively. 
Only for the intermediate scattering function showing a logarithmic decay, i.e. for the small particles and $\delta = 0.35$ and $x_\mathrm{s} = 0.01$ in the simulations and $\delta = 0.28$ and $x_\mathrm{s} = 0.01$ in the experiments, a different relation was used to describe the logarithmic decay:
\begin{equation}
f^S(q,\Delta t)=f_\mathrm{log}^S-F(q)\ln(\Delta t/\tau_\mathrm{log})
\label{log_fit}
\end{equation}

Representative fits based on Eq.~\ref{stretch_exp_fit} and on Eq.~\ref{log_fit} are reported in Fig.S4 and Fig. S5 of the supplemental material, respectively. The relaxation times $\tau^j_\mathrm{LO}$ and $\tau_\mathrm{log}$ obtained from fits to the simulation data are shown in Fig.~\ref{fit_params_1}a,b  for $\delta = 0.2$ and 0.35 as a function of $q\sigma_\mathrm{L}$ for different values of the composition $x_\mathrm{s}$. Once more we focus on a glassy sample with $\phi = 0.62$. Data for additional volume fractions $0.60\leq\phi\leq 0.63$ can be found in the supplemental material.

For $\delta = 0.2$ and all values of $x_\mathrm{s}$ the coupling of the long-time dynamics of large and small particles is evident: the relaxation times of the two species are very similar for all $q$ values. The effect of the structure factor of the large particles is visible as a series of  oscillations in the region $q\sigma_\mathrm{L} < 10$. Note that also for the small particles the oscillations follow the q-dependence of the structure factor of the large particles. This confirms that the long-time diffusion of the small particles is controlled by the structure of the large particles.  Moreover, the reduction of the relaxation times of both species indicates the melting of the glass of large particles with increasing $x_\mathrm{s}$.\\ 
Decreasing the size-disparity, i.e. at $\delta = 0.35$, for $x_\mathrm{s} = 0.01$ any conclusion about coupling is complicated by the fact that the relaxation times $\tau_\mathrm{LO}$ of the large and small particles correspond to different functional forms, due to the logarithmic decay of $f^S(q,\Delta t)$. As mentioned previously, in this case the coupling is much reduced, which is consistent with the relatively large difference of $\tau_\mathrm{LO}^\mathrm{s}$ and $\tau_\mathrm{LO}^\mathrm{L}$. For $x_\mathrm{s} = 0.05$ and 0.10 the $q$-dependence of the relaxation times is again very similar for the two species. However, compared to $\delta = 0.2$, there is a larger difference between the magnitudes of the relaxation times of the large and small particles, which is particularly pronounced (about an order of magnitude) at wavevectors corresponding to the main peak of the static structure factor. This might be attributed to a significant mutual effect of the small particles caused by the mixed caging. In a mixed cage, the small particles have a slightly higher probability to escape the cage than the large particles, since only smaller gaps are required. If the localisation is only due to large particles, as for $\delta = 0.2$ and $x_\mathrm{s} = 0.01$, the final relaxation of the small particles instead is slaved to that of the large particles. A comparison of the long-time relaxation times obtained for $\delta = 0.2$ and 0.35 evidences slower dynamics for the more moderate size ratio. The relaxation times of the small particles obtained in experiments for $x_\mathrm{s} = 0.01$  are in good qualitative agreement with those of the simulations (Fig.~\ref{fit_params_1}c). In the experiments we also determined the short-time relaxation times of small particles $\tau^S_{\mathrm{SH}}$ by fitting the short-time decay to a simple exponential dependence, indicating therefore a diffusive behaviour. Indeed the short relaxation times are compatible with a $q^{-2}$ dependence. We further notice that $\tau^S_{\mathrm{SH}}$ increases by one order of magnitude with increasing $\delta$ from 1:5 to 1:3 size ratios.

The plateau height $f_c$ also provides information about the dynamic coupling (Fig.~\ref{fit_params_2}). For $\delta = 0.2$ and $x_\mathrm{s} = 0.01$ (Fig.~\ref{fit_params_2}a) both the plateau heights for small ($f_c^S$) and large ($f_c^L$) particles display a similar $q$-dependence, which reproduces the oscillations of the structure factor of the large particles, as typically found in glassy systems. However, $f_c^S$ is much smaller than $f_c^L$, reflecting the small fraction of small particles that are trapped at long times. Experimental results are in qualitative agreement with the simulation ones (Fig.~\ref{fit_params_2}c). For the larger $x_\mathrm{s}$ values, we cannot determine $f_c^L$ since, due to glass melting, $f^L(q,\Delta t)$ displays only a single decay. 
For the small particles and larger $x_\mathrm{s}$ values, the qualitative behaviour is different from that found for $x_\mathrm{s} = 0.01$,  particularly in the region of small $q\sigma_\mathrm{L}$ where $f_c^S$ are considerably larger. This might be related to a more important effect of the interactions between small particles with increasing $x_\mathrm{s}$. The oscillations in $f_c^S$ associated to the structure factor of the large particles  also progressively decrease with increasing $x_\mathrm{s}$. This may be due to the melting of the glass of large particles or could also be caused by the interactions between small particles.

For $\delta = 0.35$ and all investigated $x_\mathrm{s}$, we find that $f_c^L$ show oscillations related to $S_\mathrm{LL}(q)$, as typical for glasses. Furthemore, we observe a weaker decrease of the plateau height value with increasing $x_\mathrm{s}$ with respect to the case $\delta = 0.2$, which could be due again to the progressive glass melting. The $f_c^S$ show similar trends for all $x_\mathrm{s}$, which are also qualitatively comparable to those of $x_\mathrm{s} = 0.05$ and 0.10 for $\delta = 0.2$. This supports the conclusion that the effect of $S_\mathrm{LL}(q)$ is decreased due to the increasingly mixed caging, in which the dynamics of the small particles is affected by the structural organization of the large particles but also of the other small particles. The experimental results for $\delta = 0.35$ and $x_\mathrm{s} = 0.01$ for the small particles are again in good qualitative agreement with simulations (Fig.~\ref{fit_params_2}c). The stretching exponents $\beta^\mathrm{s}$ are reported in the supplemental material. For most conditions, they are smaller than 1 which indicates a broad distribution of relaxation times. With increasing $q$, for  $\delta = 0.35$ and any $x_\mathrm{s}$ as well as $\delta = 0.2$ and $x_\mathrm{s} = 0.05, 0.10$, the exponents remain approximately constant or slightly decrease which suggests an increasing range of relaxation times associated to displacements on short length scales and is consistent with the general behaviour of glass-forming systems. On the contrary, in the case $\delta = 0.2$ and $x_\mathrm{s} = 0.01$, with increasing $q$,  $\beta^\mathrm{s}$ increases and approaches 1 which indicates that, at short length scales, the small particles recover a diffusive-like behaviour within the voids left by the large spheres.

In summary, the quantitative analysis of $f^L(q,\Delta t)$ and $f^S(q,\Delta t)$  indicates that the coupling of the dynamics between the two species is quite general. However, its nature depends on the composition $x_\mathrm{s}$ and size ratio $\delta$. For large size disparity and small $x_\mathrm{s}$ the long-time dynamics of the small particles is slaved to the one of the large particles. For smaller size disparity and/or larger $x_\mathrm{s}$, the coupling is also due to mixed caging, in which the dynamics of the two species mutually affect each other.

\section{Conclusions}

In this work we investigated the dynamics of model colloidal hard spheres in glassy binary mixtures, using simulations and differential dynamic microscopy experiments which allowed us to determine the dynamics  of an individual component of multicomponent samples. We focused on the connection between the dynamics of the large and small particles.  The results show different scenarios of dynamic coupling between the long-time relaxations of the two species. For the vast majority of conditions, a strong coupling between small and large particles is found. This is evidenced by a qualitative comparison of the intermediate scattering functions of the small and large particles, as well as a quantitative comparison of the relaxation times obtained by fitting a stretched exponential decay to the intermediate scattering function $f(q,\Delta t)$. However, the physical mechanism leading to the coupling depends on the size ratio $\delta$ and the composition, represented by the relative volume fraction of the small particles $x_\mathrm{s}$.  

For the largest size disparity, $\delta = 0.2$, and a small amount of small particles, $x_\mathrm{s} = 0.01$,  the short time dynamics of the two species are clearly decoupled. The small particles can diffuse within the voids left by the large particles and most of them escape the local confinement through small channels. Only a limited fraction of the small particles remains trapped for long times, but are released once the large particles moved sufficiently. This is reflected in the fact that the relaxation times and plateau heights of the small particles are affected by the structure factor of the large particles. With increasing  $x_\mathrm{s}$, this fraction becomes larger  and the coupling of the short-time dynamics of the two species becomes more evident. This might be due to the increasing importance of interactions between small particles  and the formation of mixed cages, which both contribute to the slowdown of the dynamics which hence become more similar to the dynamics of the large particles.

For larger values of $\delta$ the coupling of the long-time dynamics for $x_\mathrm{s} = 0.05$ and 0.10 is evident, but now also the short-time dynamics of the two species become more similar. This suggests a different situation compared to $\delta = 0.2$. For $\delta = 0.35$  the small particles are too large to diffuse within the voids left by the large particles. For sufficiently large $x_\mathrm{s}$ they tend to form mixed cages with the large particles, which the two species escape after similar waiting times. Interestingly, for small $x_\mathrm{s}$ the coupling becomes weaker and the $f^S(q,\Delta t)$ of the small particles shows a logarithmic decay for an intermediate range of $\delta$ that is particularly enhanced for $\delta=0.35$. As discussed in previous work \cite{nat_comm}, these dynamics arise from the confinement imposed by the large particles, which however is relaxed at longer times due to the (slow) motion of the large particles. Due to the small $x_\mathrm{s}$ mixed caging is less important and therefore the small particles are mainly trapped by large particles and are only released if the large particles move and paths open. This leads to a very broad distribution of relaxation times which depends on both the short-time and long-time relaxation of the large particles. Moreover the coupling becomes weaker, particularly at wavevectors corresponding to large-large nearest-neighbour distances.

As a general result, we observe for all size ratios an acceleration of the dynamics of the large particles with increasing the fraction of small particles. This reflects the melting of the glass due to the disruption of the cage of large particles by the intercalation of small particles. It is  more pronounced for $\delta = 0.2$, where the small particles can more easily occupy the voids left by the large particles. The glass melting is related with the transition from caging of large particles by other large particles to caging of large particles by small particles. It results in an asymmetric glass \cite{Thomas_EPL,maldonado,tanja_sm2014,tanja_review,mayer2008asymmetric}. The present results  thus suggests several different scenarios for the dynamics of binary colloidal mixtures, depending on size ratio $\delta$, volume fraction $\phi$ and mixture composition $x_\mathrm{s}$. 

\section*{Acknowledgments}
We are grateful to Andrew Schofield (University of Edinburgh) for providing the PMMA particles, Vincent Martinez (University of Edinburgh), Wilson Poon (University of Edinburgh) and Matthias Reufer (LS Instruments) for providing routines for the DDM analysis. We thank F. Sciortino and P. Tartaglia for useful discussions. T.S., S.U.E. and M.L. gratefully acknowledge funding from
the Deutsche Forschungsgemeinschaft (DFG) through the research unit FOR1394, project P2, and funding of the confocal microscope through grant INST 208/617\DH1 FUGG. M.L. acknowledges support from the project Materia Blanda Coloidal funded by Conacyt, Convocatoria de Investigacion Cientifica Basica 2014, Grant Nr. 237425 and Red Tematica de la Materia Condensada Blanda, Conacyt. JRF and EZ from ETN-COLLDENSE (H2020-MCSA-ITN-2014, Grant no. 642774).

\providecommand*{\mcitethebibliography}{\thebibliography}
\csname @ifundefined\endcsname{endmcitethebibliography}
{\let\endmcitethebibliography\endthebibliography}{}


\pagebreak
\onecolumngrid

\renewcommand{\thefigure}{S\arabic{figure}}\setcounter{figure}{0}
\section{Supplementary Online Material}
\subsection*{Additional Structural Data}

\begin{figure} [ !tbh]
      \includegraphics[angle=0, width=1.0\textwidth] {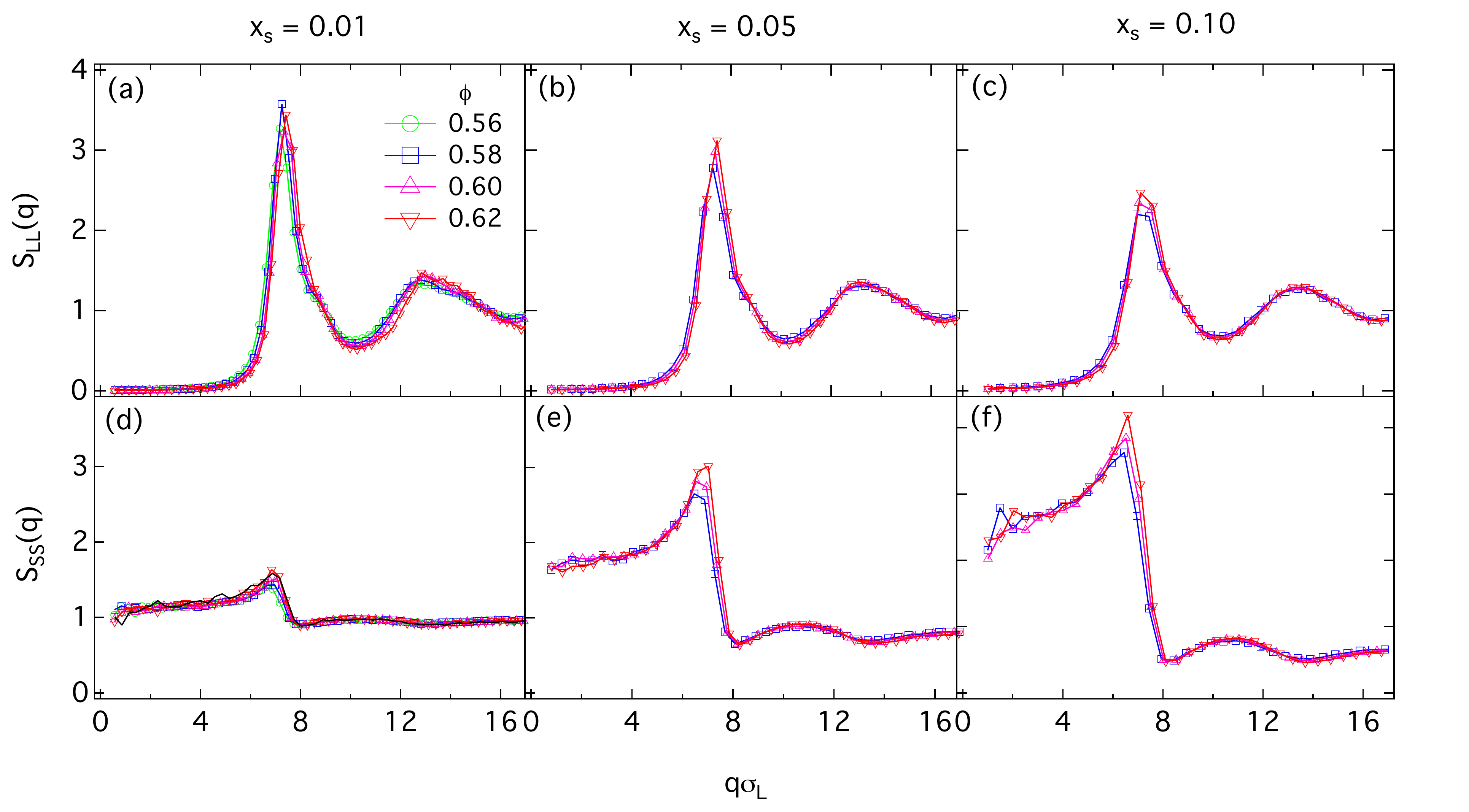}
             \caption{Static structure factors of (a,b,c) the large particles, $S_\mathrm{LL}(q)$, and (d,e,f) of the small particles, $S_\mathrm{SS}(q)$, from simulations for size ratio $\delta = 0.20$, different volume fractions $\phi$  and different fractions of small particles $x_\mathrm{s}$.}  
\label{sq_delta02}         
\end{figure}

\begin{figure} [ !tbh]
      \includegraphics[angle=0, width=1.0\textwidth] {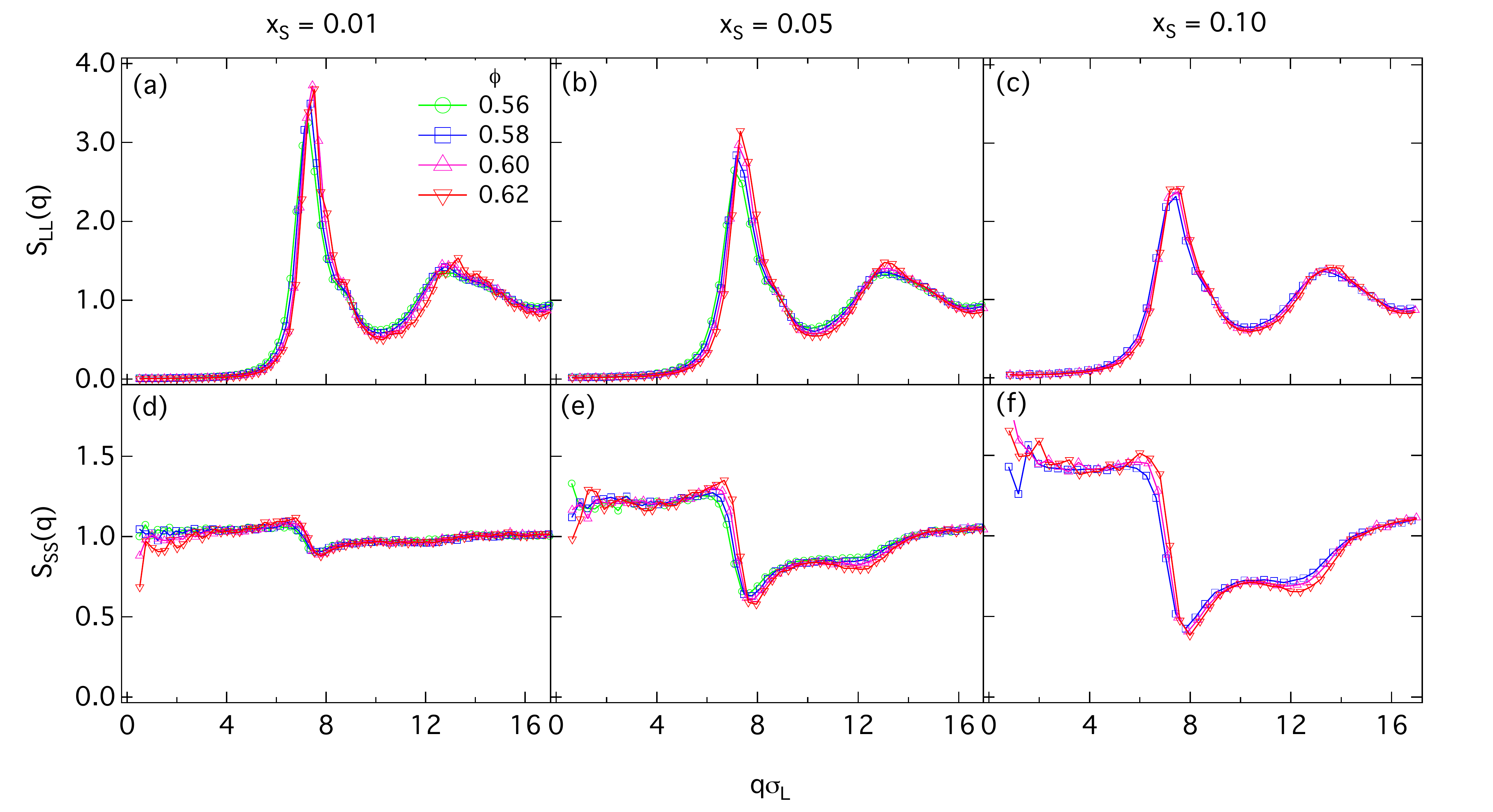}
             \caption{Static structure factors of (a,b,c) the large particles, $S_\mathrm{LL}(q)$, and (d,e,f) of the small particles, $S_\mathrm{SS}(q)$, from simulations for $\delta = 0.35$, different volume fractions $\phi$ and different fractions of small particles $x_\mathrm{s}$.}  
\label{sq_delta035}         
\end{figure}

Figs. \ref{sq_delta02} and \ref{sq_delta035} show simulation results for the partial static structure factors of large and small particles, $S_\mathrm{LL}(q)$ and $S_\mathrm{SS}(q)$, respectively, for different volume fractions $\phi$, size ratios $\delta$ and compositions $x_\mathrm{s}$. These data complement the data in Fig.1 of the main manuscript for $\phi = 0.62$. As it can be seen, changes in volume fraction do not affect the qualitative features of the structure factors presented in Fig.1 of the main manuscript, they mainly affect the height of the peaks. 

\begin{figure} [ !tbh]
      \includegraphics[angle=0, width=0.5\textwidth] {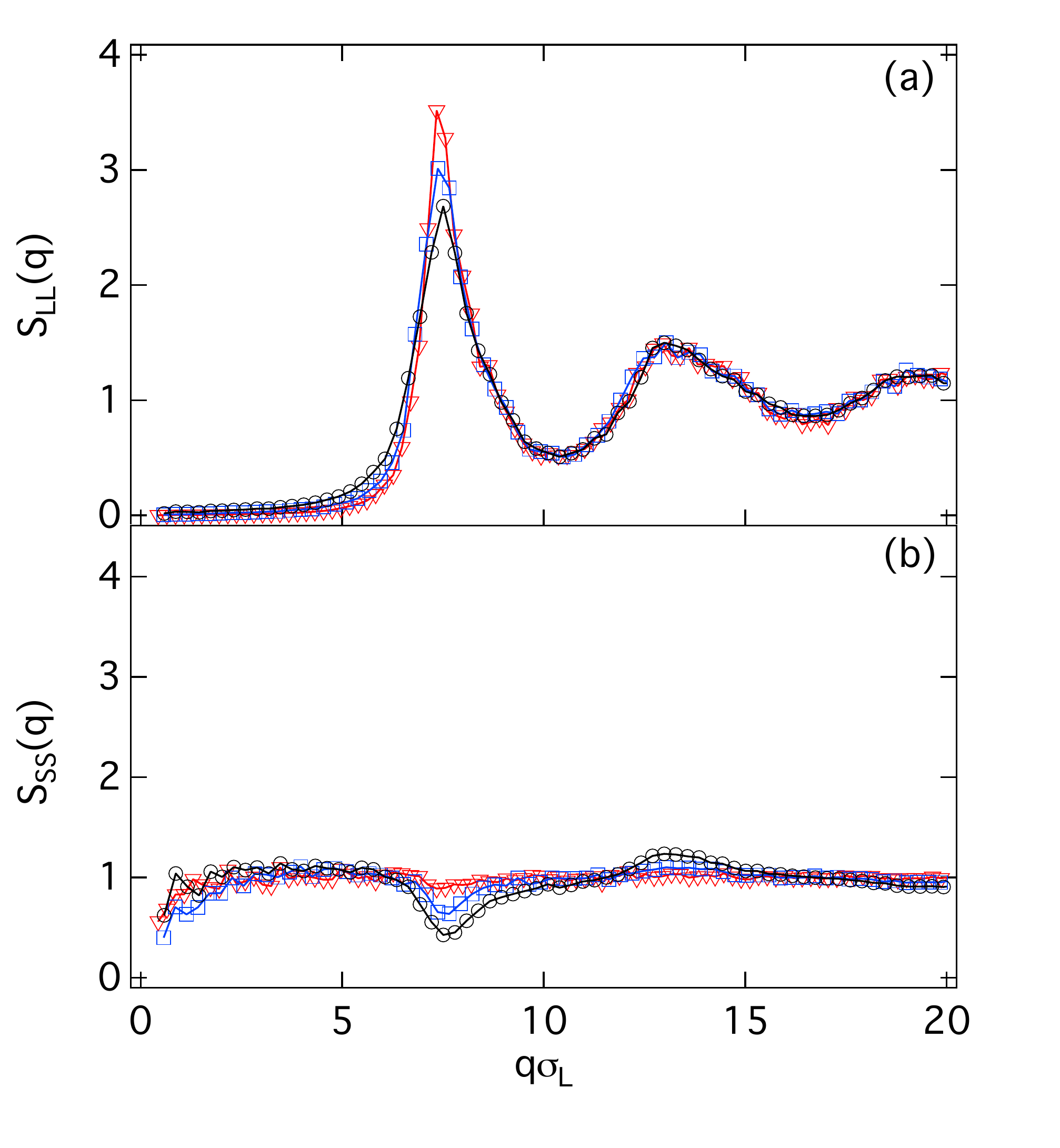}
             \caption{Static structure factors of (a) the large  particles, $S_\mathrm{LL}(q)$, and (b) small particles, $S_\mathrm{SS}(q)$ from simulations for size ratio $\delta = 0.5$, volume fraction $\phi = 0.62$ and fractions of small particles $x_\mathrm{s}= 0.01$ (triangles), 0.05 (squares) and 0.10 (circles).} 
\label{sq_delta05}         
\end{figure}

In addition, Fig. \ref{sq_delta05} completes the data reported in Fig.1 of the manuscript with the partial static structure factors of large and small particles for $\delta=0.5$. While $S_{LL}(q)$ has a very similar behavior to the other size ratios reported in the manuscript, showing only a decrease of the peaks with increasing $x_\mathrm{s}$, $S_{SS}(q)$ displays a behavior that is distinct from the other two size ratios:   the usual dependence  for $x_\mathrm{s}=0.01$ is followed by the occurrence of a dip (rather than a peak) for $\delta = 0.05$ and 0.10 which corresponds to the first peak of $S_{LL}(q)$ of the large particles. This indicates the strong coupling between large and small particles at this size ratio.

\section*{Additional Dynamical Data}

\begin{figure*} [!tbh]
   \centering
       \includegraphics[angle=0, width=1.0\textwidth] {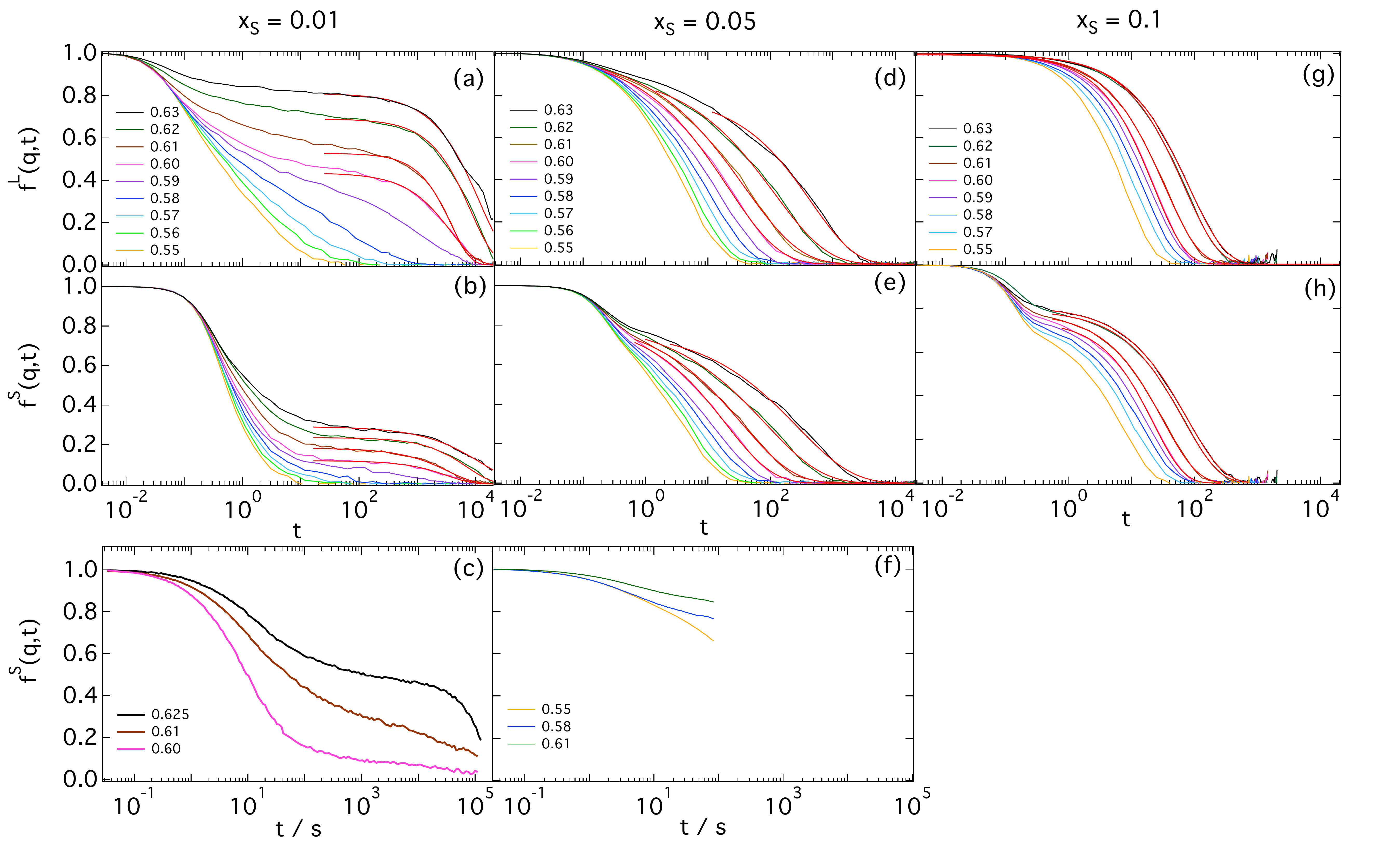}
         \caption{Intermediate scattering functions $f(q,t)$ for $\delta = 0.2$, $q\sigma_l = 3.5$, different volume fractions $\phi$ and different values of $x_\mathrm{s}$. Top graphs (a,d,g) show results of simulations for the large particles, mid graphs (b,e,h) for the small particles. Red lines in (a,b,d,e,g,h) represent fits of the long time relaxation of $f(q,t)$ according to Eq.6 of the main manuscript. Bottom graphs (c,f) show corresponding experimental results for the small particles for $\delta = 0.18$.} 
\label{phi_dep_1}          
\end{figure*}
\begin{figure*} [ !tb]
      \includegraphics[angle=0, width=1.0\textwidth] {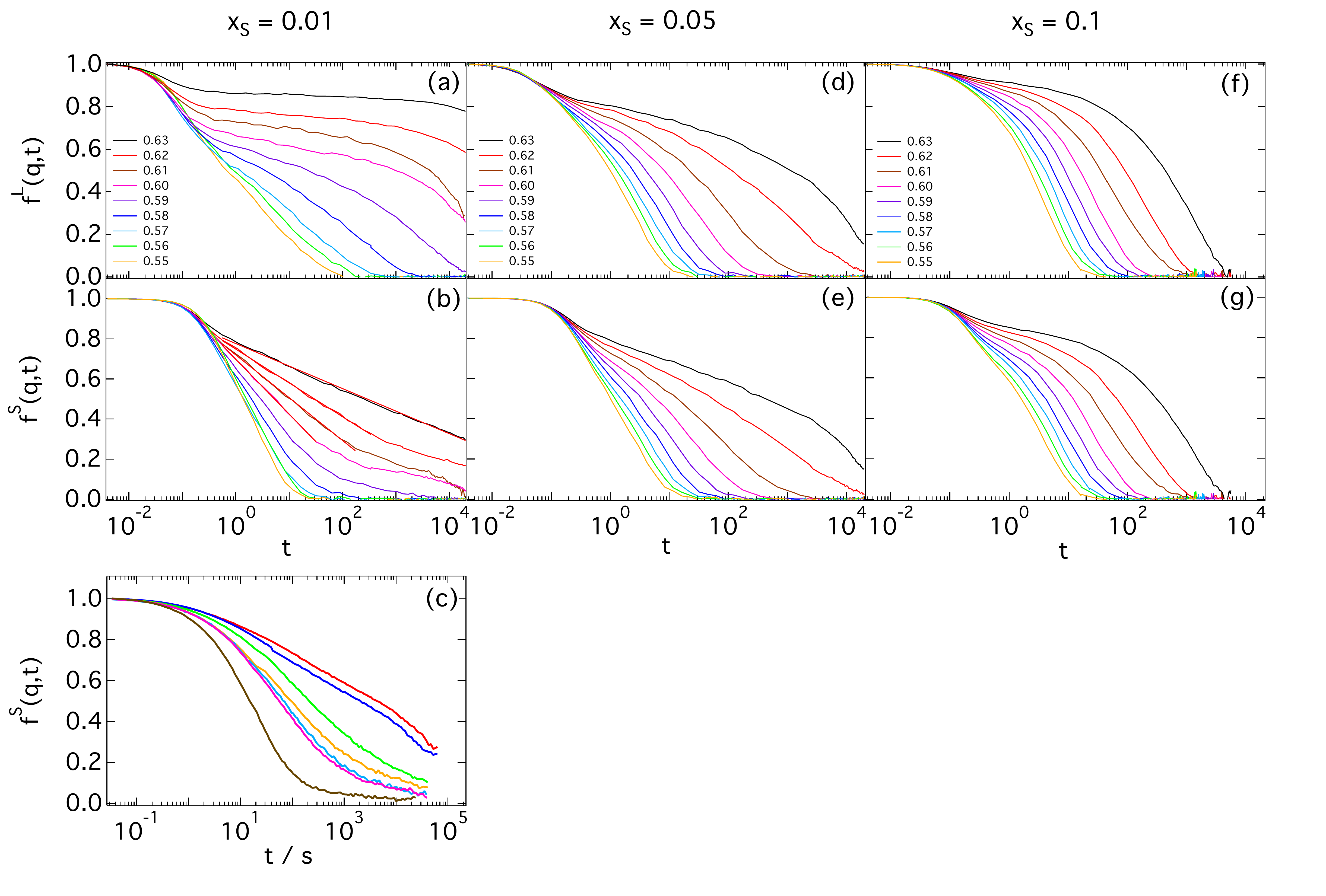}
             \caption{Intermediate scattering functions $f(q,t)$ for $\delta = 0.35$, $q\sigma_l = 3.5$, different volume fractions $\phi$ and different values of $x_\mathrm{s}$. Top graphs (a,d,g) show results of simulations for the large particles, mid graphs (b,e,h) for the small particles. Red lines in (b) represent fits of the logarithmic relaxation of $f(q,t)$ according to Eq.7 of the main manuscript. Bottom graphs show corresponding experimental results for the small particles for $\delta = 0.28$.}  
\label{phi_dep_2}         
\end{figure*}

Figs. \ref{phi_dep_1} and \ref{phi_dep_2} present simulation results for the collective intermediate scattering functions $f(q,t)$ of large and small particles obtained for different volume fractions $\phi$, compositions $x_\mathrm{s}$ and size ratios $\delta$. These data complement those of Fig.2 of the main manuscript for $\phi = 0.62$. As it can be seen, the reduction in total volume fraction $\phi$ results in a progressive acceleration of the decay of $f(q,t)$, associated with the dilution of the system. It is interesting to note in Fig.\ref{phi_dep_2} that for $\delta = 0.35$ the anomalous logarithmic relaxation for the small particles is observed over the whole time window only for the largest total volume fraction, $\phi = 0.63$, while for the smaller total volume fractions it reduces to a shorter time interval.

\begin{figure*} [!tb]
\centering
     \includegraphics[angle=0, width=1.0\textwidth] {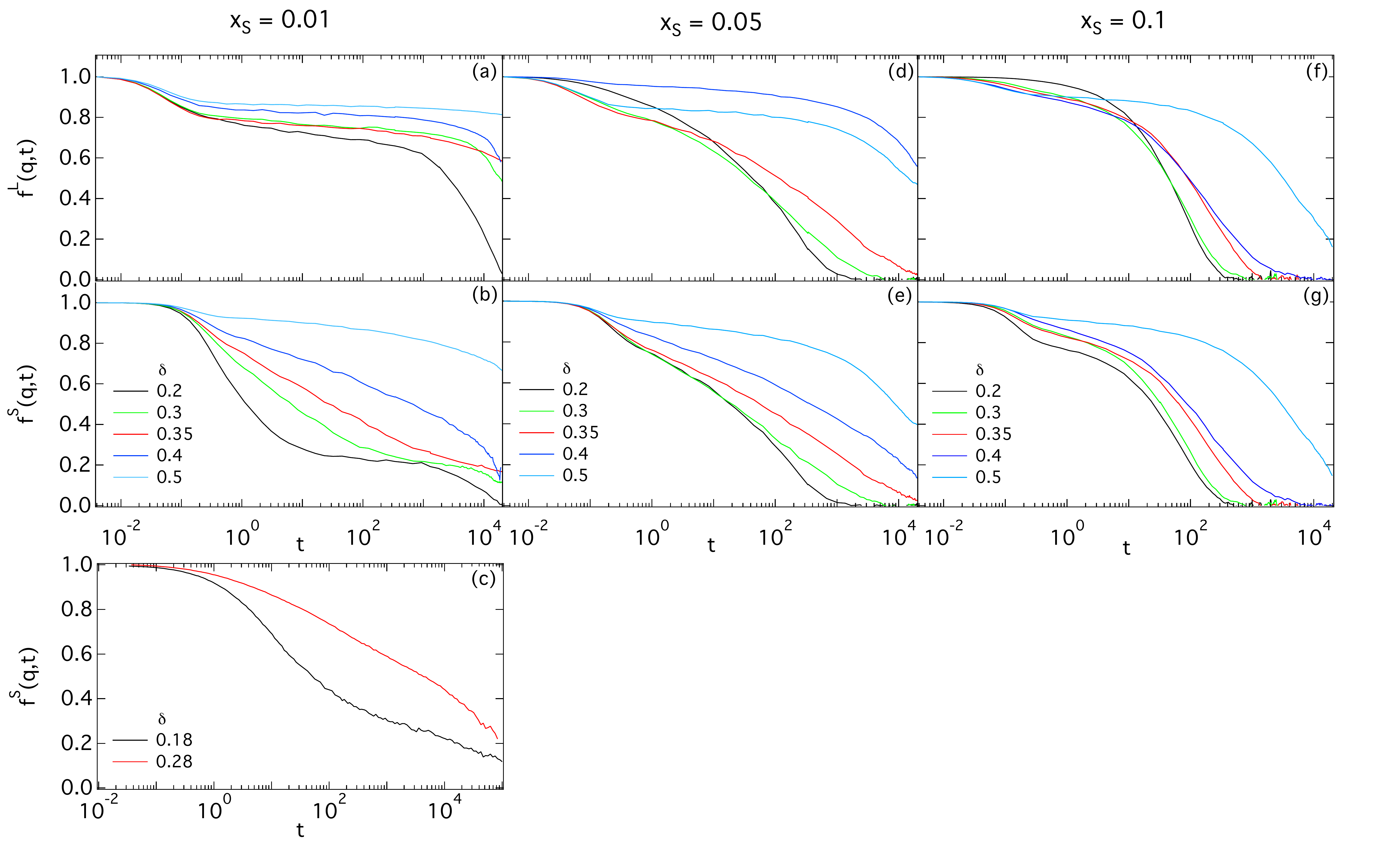}
\caption{Intermediate scattering functions $f(q,t)$ for different $\delta$, $q\sigma_l = 3.5$ and different values of $x_\mathrm{s}$:  Top graphs (a,d,f) show results of simulations for the large spheres, $\phi = 0.62$, mid graphs (b,e,g) for the small particles, $\phi = 0.62$. Bottom graph (c) shows corresponding experimental results for small particles and $\phi = 0.61$.}         
\label{fig:sizeratio}
\end{figure*}

Fig. \ref{fig:sizeratio} shows the collective intermediate scattering functions $f(q,t)$ obtained for $\phi = 0.62$, $q\sigma_l = 3.5$, different compositions $x_\mathrm{s}$ and size ratios $\delta$. These data complement those presented in Fig.4 of the main manuscript. It can be seen that with decreasing size disparity (increasing $\delta$) the dynamics of the small and large particles become increasingly comparable at short and long times, indicating the formation of mixed cages.

\section*{Fit parameters}

\begin{figure*} [!tb]
\centering
     \includegraphics[angle=0, width=0.95\textwidth] {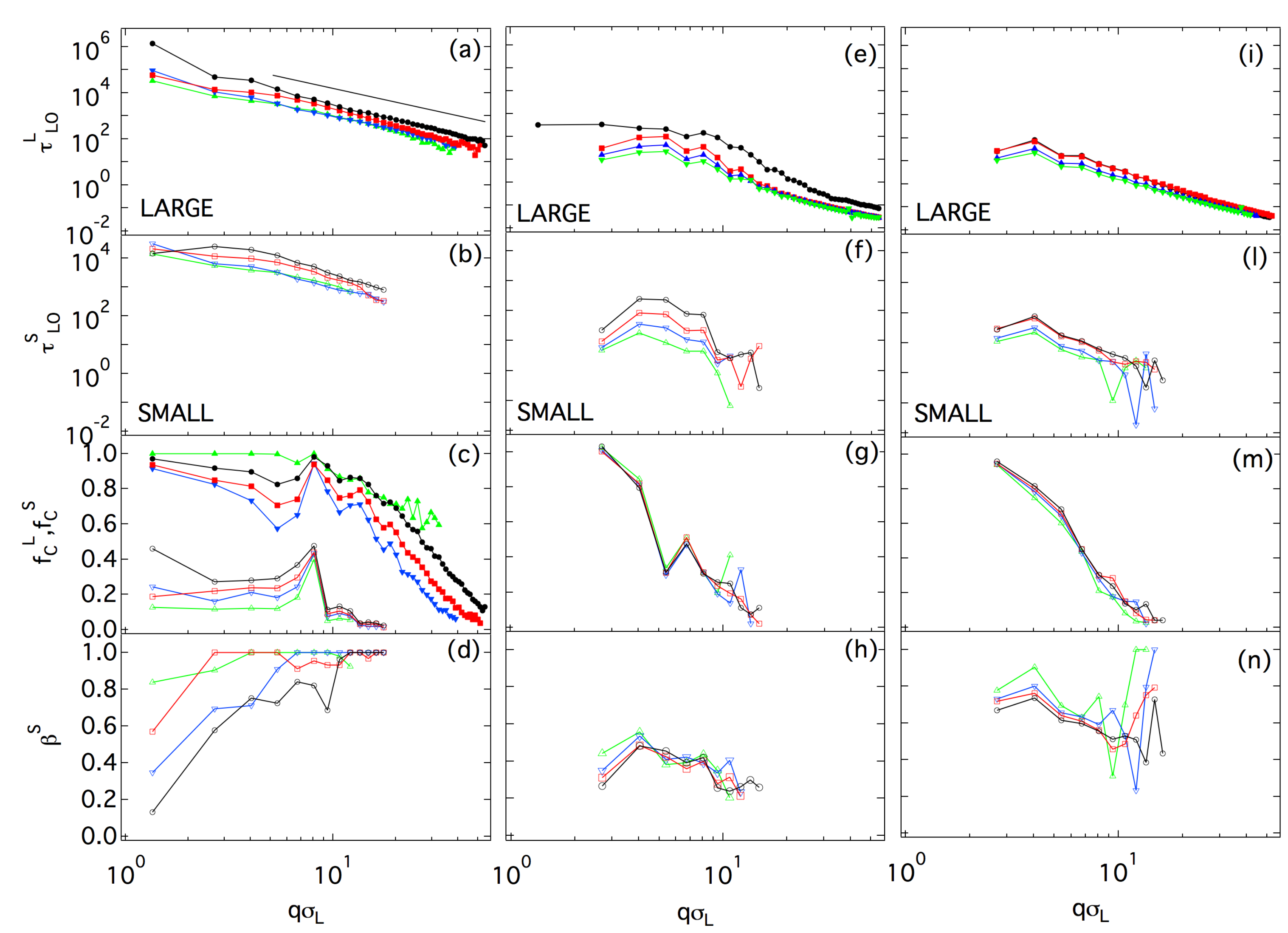}
\caption{Parameters obtained by fitting Eq. 6 of the main manuscript to $f(q,t)$ from simulations for $\delta = 0.2$, $\phi = 0.60$ (\textcolor{green}{$\triangle$}), 0.61 (\textcolor{blue}{$\triangledown$}), 0.62 (\textcolor{red}{$\Box$}) and 0.63 ({\Large $\circ$}) and $x_\mathrm{s} = 0.01$ (left column), 0.05 (mid column) and 0.1 (right column), as a function of  $q\sigma_L$.  (a), (e), (i):  Long-time relaxation times for the small particles, $\tau^S_\mathrm{LO}$ (open symbols). (b), (f), (l):  Long-time relaxation times for the large particles, $\tau^L_\mathrm{LO}$ (full symbols). (c), (g), (m): Plateau height for the large, $f_{c}^{L}$ (full symbols, when present), and small, $f_{c}^{S}$ (open symbols), particles. (d), (h), (n): Stretching exponent $\beta^S$ obtained from the fits to the long-time relaxation of the small particles.}
\label{fit_params_1}         
\end{figure*}

\begin{figure} [!tb]
\centering
     \includegraphics[angle=0, width=0.95\textwidth] {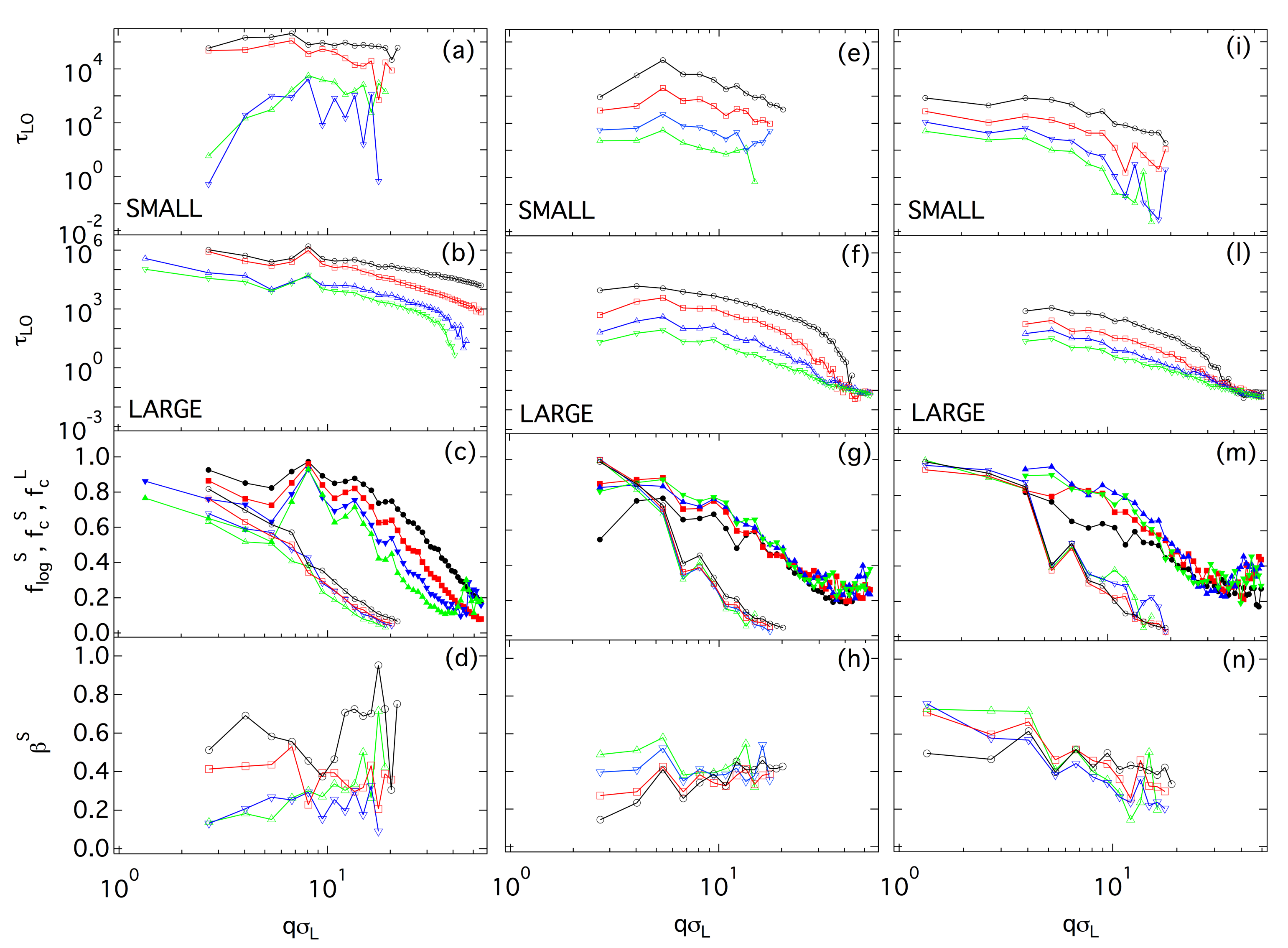}
\caption{Parameters obtained by fitting Eq. 6 of the main manuscript to $f(q,t)$ from simulations for $\delta = 0.35$, $\phi = 0.62$ and $x_\mathrm{s} = 0.01$ ($\Box$), 0.05 (\textcolor{red}{\Large $\circ$}), 0.10 (\textcolor{blue}{$\triangle$}), as a function of  $q\sigma_L$.  (a), (e), (i):  Long-time relaxation times for the small particles, $\tau^S_\mathrm{LO}$ (open symbols). (b), (f), (l):  Long-time relaxation times for the large particles, $\tau^L_\mathrm{LO}$ (full symbols).   (c), (g), (m): Plateau height for the large, $f_{c}^{L}$ (full symbols, when present), and small, $f_{c}^{S}$ (open symbols), particles. For $x_\mathrm{s} = 0.01$ we report the logarithmic non-ergodicity parameter $f_\mathrm{log}^S$ as defined in Eq.7 of the main manuscript. (d), (h), (n): Stretching exponent $\beta^S$ obtained from the fits to the long-time relaxation of the small particles.}         
\label{fit_params_2} 
\end{figure}

\begin{figure*} [!tb]
\centering
     \includegraphics[angle=0, width=0.7\textwidth] {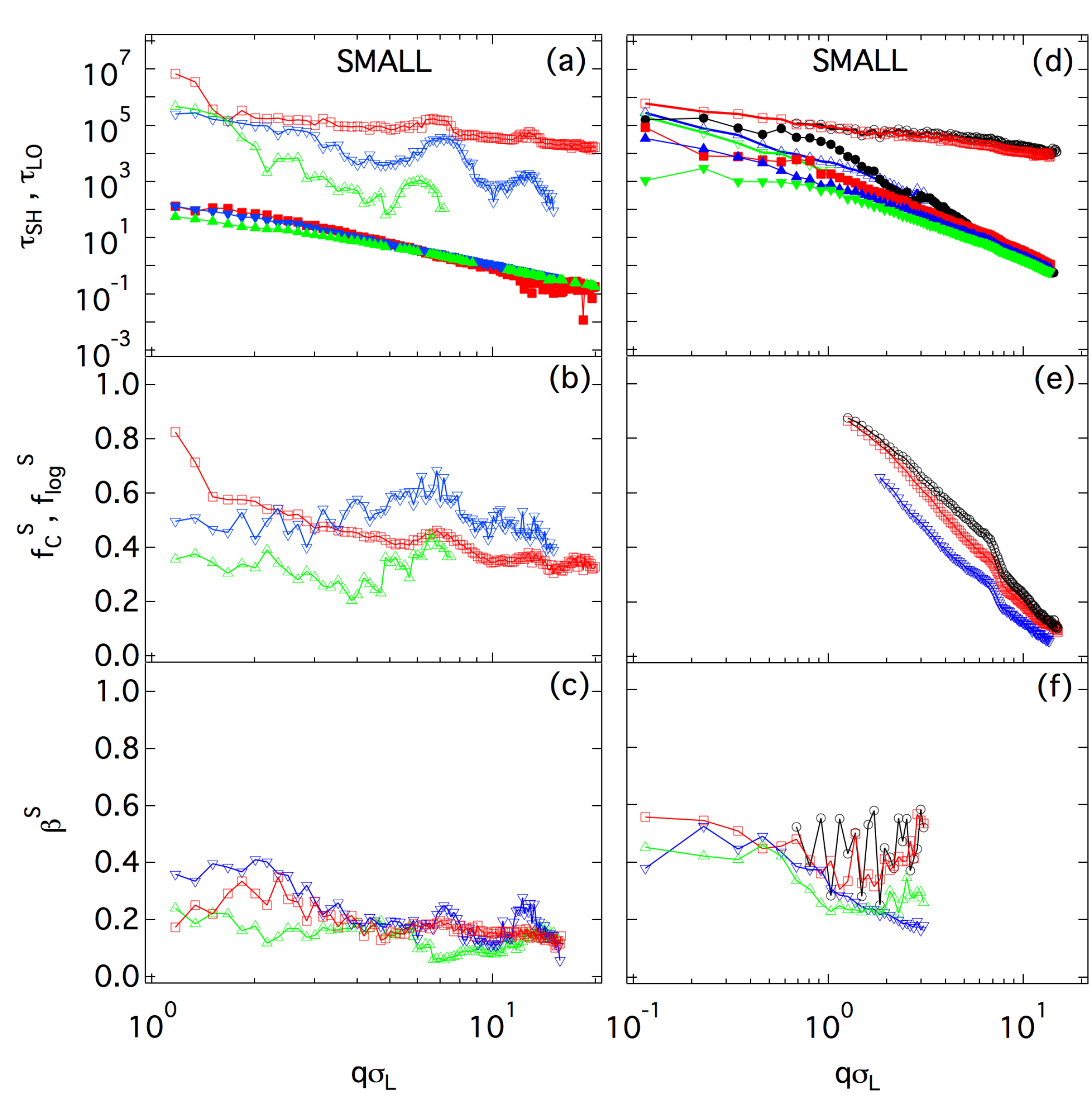}
\caption{Parameters obtained by fitting Eq.6 to $f(q,t)$ from experiments for $x_\mathrm{s} = 0.01$, $\delta = 0.18$ and $\phi = 0.60$ (\textcolor{green}{$\triangle$}), 0.61 (\textcolor{blue}{$\triangledown$}), 0.625 (\textcolor{red}{$\Box$}) (Left)  and $\delta = 0.28$ $\phi = 0.58$  (\textcolor{green}{$\triangle$}), 0.59 (\textcolor{blue}{$\triangledown$}), 0.60 (\textcolor{red}{$\Box$}) 0.61 ({\Large $\circ$}) (Right).  (a) ,(d) Long and short relaxation times of the small particles, $\tau_\mathrm{LO}$ (open symbols) and $\tau_\mathrm{SH}$ (full symbols), respectively. (b), (e): Plateau height $f_c^S$ (b) and $f_\mathrm{log}^S$ (e), where the last is defined in Eq.7 of the main manuscript. (c), (f) Stretching exponents obtained $\beta^S$ obtained from the fits to the long-time relaxation of the small particles.}         
\label{fit_params_exp} 
\end{figure*}

Fit parameters obtained by fitting a stretched exponential function according to Eq.6 of the main article to $f(q,t)$ of the large and small particles from simulations (Figs. \ref{fit_params_1} and \ref{fit_params_2}) and experiments (Fig. \ref{fit_params_exp}). For the small particles and $\delta = 0.35$ and $x_\mathrm{s} = 0.01$ conditions, Eq.7 of the main manuscript was used and, in agreement with previous work\cite{meyer_zaccarelli} $\tau_\mathrm{log} = 5$ was fixed. These data complement those reported in Figs. 5 and 6 of the main article. The coupling of the dynamics of the two species, already discussed in the main article for $\phi = 0.62$, is present also for other total volume fractions $\phi$, as can be seen by comparing the long-time relaxation times of the small and large particles, $\tau_\mathrm{LO}^S$ and $\tau_\mathrm{LO}^L$, respectively, as well as the plateau heights $f_c^S$ and $f_c^L$, respectively. The small values of the stretching exponents found for essentially all samples evidence the broad distribution of relaxation times of the small particles.

\end{document}